\documentclass[twocolumn,tighten,times]{aastex62}
\usepackage{color}
\usepackage{multirow}
\usepackage{chngcntr}
\usepackage[utf8x]{inputenc}
\usepackage{perpage}
\MakePerPage{footnote}
\usepackage{lipsum}
\usepackage[graphicx]{realboxes}
\graphicspath{{./}{figures/}}

\received{\today}
\revised{tomorrow}
\accepted{the day after tomorrow}

\submitjournal{ApJ}

\shorttitle{UDGs Morphology}
\shortauthors{Rong et al.}

\begin{document}

\title{Intrinsic Morphology of Ultra-diffuse Galaxies}

\correspondingauthor{Yu Rong}
\email{rongyuastrophysics@gmail.com}

\author[0000-0002-2204-6558]{Yu Rong}
\altaffiliation{FONDECYT Postdoctoral Fellow}
\affiliation{Institute of Astrophysics, Pontificia Universidad Cat\'olica de Chile, Av.~Vicu\~na Mackenna 4860, 7820436 Macul, Santiago, Chile}

\author[0000-0002-4002-861X]{Xiao-Yu Dong}
\affiliation{Department of Physics and Astronomy, California State University, Northridge, California 91330, USA}

\author[0000-0003-0350-7061]{Thomas H.~Puzia}
\affiliation{Institute of Astrophysics, Pontificia Universidad Cat\'olica de Chile, Av.~Vicu\~na Mackenna 4860, 7820436 Macul, Santiago, Chile}

\author[0000-0002-8835-0739]{Gaspar Galaz}
\affiliation{Institute of Astrophysics, Pontificia Universidad Cat\'olica de Chile, Av.~Vicu\~na Mackenna 4860, 7820436 Macul, Santiago, Chile}

\author[0000-0003-4945-0056]{Ruben S\'anchez-Janssen}
\affiliation{STFC UK Astronomy Technology Centre, Royal Observatory, Blackford Hill, Edinburgh, EH9 3HJ, UK}

\author{Tianwen Cao}
\affiliation{Chinese Academy of Sciences South America Center for Astronomy, National Astronomical Observatories, Chinese 
Academy of Sciences, Beijing 100012, China}
\affiliation{Key Laboratory of Optical Astronomy, National Astronomical Observatories, Chinese Academy of Sciences, Beijing 100101, China}
\affiliation{School of Astronomy and Space Science, University of Chinese Academy of Sciences, Beijing 100049, China}
\affiliation{Institute of Astrophysics, Pontificia Universidad Cat\'olica de Chile, Av.~Vicu\~na Mackenna 4860, 7820436 Macul, Santiago, Chile}

\author{Remco F. J. van der Burg}
\affiliation{European Southern Observatory, Karl-Schwarzschild-Str. 2, 85748, Garching, Germany}

\author{Crist\'obal Sif\'on}
\affiliation{Instituto de F\'isica, Pontificia Universidad Cat\'olica de Valpara\'iso, Casilla 4059, Valpara\'iso, Chile}
\affiliation{Department of Astrophysical Sciences, Peyton Hall, Princeton University, Princeton, NJ 08544, USA}

\author{Pavel E. Mancera Pi\~na}
\affiliation{Kapteyn Astronomical Institute, University of Groningen, Lanleven 12, 9747 AD, Groningen, The Netherlands}
\affiliation{ASTRON, the Netherlands Institute for Radio Astronomy, Postbus 2, 7990 AA, Dwingeloo, The Netherlands}

\author{Mora Marcelo}
\affiliation{Institute of Astrophysics, Pontificia Universidad Cat\'olica de Chile, Av.~Vicu\~na Mackenna 4860, 7820436 Macul, Santiago, Chile}

\author{Giuseppe D'Ago}
\affiliation{Institute of Astrophysics, Pontificia Universidad Cat\'olica de Chile, Av.~Vicu\~na Mackenna 4860, 7820436 Macul, Santiago, Chile}

\author{Hong-Xin Zhang}
\affiliation{CAS Key Laboratory for Research in Galaxies and Cosmology, Department of Astronomy, University of Science and Technology of China, China}
\affiliation{School of Astronomy and Space Sciences, University of Science and Technology of China, Hefei, 230026, China}

\author[0000-0002-2368-6469]{Evelyn J. Johnston}
\altaffiliation{FONDECYT Postdoctoral Fellow}
\affiliation{Institute of Astrophysics, Pontificia Universidad Cat\'olica de Chile, Av.~Vicu\~na Mackenna 4860, 7820436 Macul, Santiago, Chile}

\author[0000-0001-8654-0101]{Paul~Eigenthaler}
\affiliation{Institute of Astrophysics, Pontificia Universidad Cat\'olica de Chile, Av.~Vicu\~na Mackenna 4860, 7820436 Macul, Santiago, Chile}
\affiliation{Astro-Engineering Center of the Catholic University (AIUC), Av. Vicu\~na Mackenna 4860 , 7,820,436 Macul, Santiago, Chile}
\affiliation{Max-Planck-Institut f\"ur Astronomie, K\"onigstuhl 17, D-69117 Heidelberg, Germany}


\begin{abstract}
	
	With the published data of apparent axis ratios for 1109 ultra-diffuse galaxies (UDGs) located in 17 low-redshift ($z\sim 0.020\--0.063$) galaxy clusters and 84 UDGs in 2 intermediate-redshift ($z\sim 0.308\--0.348$) clusters, we take advantage of a Markov Chain Monte Carlo approach and assume a triaxial model to investigate the intrinsic morphologies of UDGs. In contrast to the conclusion of \cite{Burkert17}, i.e., the underlying shapes of UDGs are purely prolate ($C=B<A$), we find that the data favor the oblate-triaxial models (i.e., thick disks with $C<B\lesssim A$) over the nearly prolate ones. We also find that the intrinsic morphologies of UDGs are related to their stellar masses/luminosities, environments, and redshifts. First, the more-luminous UDGs have puffier morphologies compared with the less-luminous counterparts; the UDG morphologic dependence on luminosity is distinct from that of the typical quiescent dwarf ellipticals (dEs) and dwarf spheroidals (dSphs); in this sense, UDGs may not be simply treated as an extension of the dE/dSph class with similar evolutionary histories; they may differ not only in size. Second, the UDGs with smaller cluster-centric distances are more puffed-up, compared with the counterparts with larger cluster-centric distances; in combination with the UDG thickness dependence on luminosity, the puffier morphologies of UDGs with high luminosities or located in the denser environments are very likely to be attributed to tidal interactions with massive galaxies. Third, we find that the intermediate-redshift UDGs are more flattened, compared with the low-redshift counterparts, which plausibly suggests a `disky' origin for the high-redshift, newly-born UDGs.

\end{abstract}

\keywords{galaxies: dwarf --- galaxies: stellar content --- galaxies: photometry --- galaxies: evolution --- galaxies: interactions --- methods: statistical}


\section{Introduction} \label{sec:intro}

The investigations of the properties and formation of ultra-diffuse galaxies \citep[UDGs;][]{vanDokkum15} with Milky-Way sizes but luminosities of typical dwarfs, which were firstly reported about three decades ago \citep{Sandage84,Impey88,Caldwell87,Conselice03}, have not reached a clear consensus. The high-redshift strong feedback model agrees that UDGs may be failed $L^*$ galaxies embedded in massive halos but ceased their in-situ star formation in the early Universe \citep[e.g.,][]{Yozin15}; the observations for the largest UDG in the Coma cluster, DF~44, also support UDGs to be hosted in massive halos with total masses of $M_{\rm{h}}\sim 10^{12}\ M_{\odot}$ \citep{vanDokkum16}. However, the semi-analytic galaxy formation models and some hydrodynamical simulations prefer UDGs to populate the relatively lower mass halos $M_{\rm{h}}\sim 10^{9}\--10^{11}\ M_{\odot}$, originated from the high-specific-angular-momentum of their host halos or outflows \citep[e.g.,][]{Amorisco16,Rong17,DiCintio17}; the halo masses for the UDGs located in galaxy clusters, estimated from the empirical relation between masses of member globular cluster system and their parent halo \citep[e.g.][]{Prole19,Peng16,Beasley16a,Beasley16b} as well as gravitational lensing technique \citep{Sifon18}, support that UDGs may be genuine dwarfs; the morphological and structural classifications and stellar population properties of many UDGs were found to be similar to those of the dE/dSphs, also suggesting UDGs and dE/dSphs are taken from the same population and differ only in size \citep{Conselice18,Wittmann17}; the H{\,\sc{I}} detections for the UDGs located in the low-density environments reveal the relatively higher specific angular momenta of these field UDGs, compared with those of the typical dwarf counterparts \citep{Leisman17}. Other possible origins, including tidal interaction and supernova-energy injection, etc. \citep{Conselice18,Carleton19,Ogiya18,Jiang18,Martin19,Liao19}, are also plausibly supported by the controversial photometric tidal/interaction evidence \citep[e.g.,][]{Greco18,Bennet18,Muller19} as well as spectroscopic results \citep{Chilingarian19,Martin-Navarro19,Struble18}; particularly, the recent findings of lack of dark matter in two member UDGs in the NGC~1052 group, NGC~1052-DF2 and -DF4, possibly suggest an outstanding role of tidal stripping in UDG evolution \citep{vanDokkum18,vanDokkum19,Danieli19,Ogiya18}.

For the member UDGs in the low-density and high-density environments, they primarily populate the blue cloud and red sequence in the color-magnitude diagram, respectively \citep[e.g.,][]{vanderBurg16,Janssens17,Roman17a,Roman17b,Venhola17,Spekkens18,Mihos15,Lee17,Trujillo17,Rong20b}; their morphologies are also distinct: the former ones are mostly irregular, while the latter ones usually have elliptical appearances \citep[e.g.,][]{Leisman17,Trujillo17,Yagi16,Roman17a,Eigenthaler18,Conselice18}. A large fraction of UDGs in galaxy clusters exhibit unresolved nuclear star clusters \citep[e.g.,][]{Yagi16,Eigenthaler18,Mihos15}.
Several UDGs show clear evidence for association with tidal material and interaction with a larger galaxy halo \citep[e.g.,][]{Toloba16,Bennet18}.
These photometric evidences suggest the diverse morphologies of UDG populations and plausibly imply the evolution of UDG intrinsic morphologies with redshifts and environments, which further provide a clue to the formation and evolution of UDGs.

According to the distribution of the apparent axis ratios $q=b/a$ of the Coma UDGs, in particular, the absence of UDGs with $q>0.9$, \cite{Burkert17} claimed that the on-average intrinsic shapes of the cluster UDGs are more likely to be purely-prolate (i.e., the three intrinsic axes of UDGs satisfy $C=B<A$), compared with a purely-oblate disk model (i.e., $C<B=A$); the strong radial alignment signals{\footnote{there is also literature alternatively suggests no UDG radial alignment in some galaxy clusters \citep{Pina19,Rong19}, but possible UDG primordial alignment \citep{Rong20}.}} of cluster UDGs \citep[e.g.,][]{vanderBurg17,Yagi16} may also prefer a prolate model. However, it is worth to note that a more reasonable diagnostic for the underlying morphologies of UDGs is to assume a prevalent triaxial ($C\leq B\leq A$) model, rather than to simply choose between the purely-prolate and purely-oblate models. Specifically, the sharply reduced number of Coma UDGs with $q>0.9$ in the $q$ distribution can also be well explained by an oblate-triaxial model \citep[e.g.,][cf. section 4.3.3 and Fig.~4.36]{Binney98}, except for the purely-prolate model.

Therefore, the three-dimensional (3D) morphologies of UDGs should be carefully studied again with a triaxial model. We aim to analyze the possible evolution of UDG morphologies from, e.g., low-density to high-density environments, high-redshifts to low-redshifts, and low-mass to high-mass, etc. In section~\ref{sec:2}, we introduce the UDG samples studied in this work, and show the distributions of their apparent axis ratios. In section~\ref{sec:3}, we investigate the intrinsic shapes of UDGs by assuming a triaxial model, and study the possible morphology evolution of UDGs. We summarize our results in section~\ref{sec:4}.

\section{UDG data}\label{sec:2}

\subsection{UDG samples}\label{sec:2.1}

\begin{table}[!] 
\begin{tabular}{@{}lccccc@{}}
\hline
\hline
Clusters & $z$ & $R_{200}$ (Mpc) & $N_{\rm{inner}}$ & $N_{\rm{middle}}$ & $N_{\rm{outer}}$ \\
\hline
Coma$^{a}$ & 0.023 & 2.6$^{*}$ & 204 & 124 & 0 \\
R1204$^{b}$ & 0.020 & 0.6  & 7 & 15 & 17 \\
A779$^{b}$ & 0.023 & 0.7  & 7 & 17 & 7 \\
R1223$^{b}$ & 0.026 & 0.6  & 10 & 7 & 20 \\
MKW4S$^{b}$ & 0.027 & 0.6  & 5 & 9 & 28 \\
R1714$^{b}$ & 0.028 & 0.4  & 2 & 6 & 35 \\
A2634$^{b}$ & 0.031 & 1.3  & 51 & 61 & 8 \\
A1177$^{b}$ & 0.032 & 0.7  & 5 & 9 & 25 \\
A1314$^{b}$ & 0.033 & 0.9  & 16 & 20 & 55 \\
A119$^{c}$ & 0.044 & 1.9  & 38 & 18 & 0 \\
MKW3S$^{c}$ & 0.044 & 1.2  & 8 & 10 & 0 \\
A85$^{c}$ & 0.055 & 2.0  & 37 & 30 & 0 \\
A780$^{c}$ & 0.055 & 1.7  & 11 & 23 & 0 \\
A133$^{c}$ & 0.056 & 1.7  & 27 & 27 & 0 \\
A1991$^{c}$ & 0.059 & 1.2  & 17 & 8 & 0 \\
A1781$^{c}$ & 0.062 & 0.9  & 4 & 12 & 0 \\
A1795$^{c}$ & 0.063 & 1.6  & 22 & 47 & 0 \\
A2744$^{d}$ & 0.308 & 2.4  & 26 & 13 & 0 \\
AS1063$^{d}$ & 0.348 & 2.5  & 33 & 6 & 6 \\
\hline
\hline
\end{tabular}
\caption{Information of the parent galaxy clusters/groups where the UDG samples located. 
Col. (1): cluster name; Col. (2): redshift; Col. (3): virial radius $R_{200}$ (Mpc); 
Col. (4)\--(6): numbers of UDGs in $R\leq 0.5R_{200}$, $0.5R_{200}<R\leq R_{200}$, and $R>R_{200}$, respectively. $^a$ \cite{Yagi16}; $^b$ \cite{Pina19}; $^c$ \cite{vanderBurg16}; $^d$ \cite{Lee17}; $^*$ \cite{Brilenkov15}. 
}
\label{UDG_samples}
\end{table}

The UDG samples used in this work are gathered from the previous literature, located in 17 low-redshift (low-$z$; $z\sim 0.020\--0.063$) clusters/groups and 2 intermediate-redshift (intermediate-$z$; $z\sim 0.308\--0.348$) clusters, as listed in Table~\ref{UDG_samples}. Almost all of these UDGs follow the red sequence in the color-magnitude diagram.

Sample~1: the publicly available {\footnote{http://vizier.cfa.harvard.edu/viz-bin/VizieR?-source=J/ApJS/225/11}} Coma UDG sample reported by \cite{Yagi16}.
These UDGs are distributed within $R_{\rm{200}}$ ($R_{200}$ is the radius within which the mean cluster density is 200 times of the critical density) of Coma, with $r$-band{\footnote{The original magnitude and surface brightness values of UDGs in \cite{Yagi16} are in the $R$-band; in order to compare the surface brightness and magnitudes of the Coma UDGs with those of the other UDG samples, we convert the $R$-band surface brightnesses and magnitudes to $r$-band properties with $R-r=\Sigma^{7}_{k=0}c_k(r-i)^k\sim 0.08$~mag \citep{Yagi16}, where the colors of the cluster UDGs approximately are $r-i\sim 0.25$ \citep[e.g.,][]{Rong17}, and constants $c_k$ are obtained from Table~2 in \cite{Yagi16}. Indeed, the $R$ and $r$-band magnitudes/surface brightness levels only show a marginal difference.}} absolute magnitudes of $-17<M_r<-9$~mag, effective radii $r_{\rm{e}}>1.5$~kpc, as well as mean surface brightness within $r_{\rm{e}}$, $\langle\mu_{\rm{e}}(r)\rangle$, between 24 and 27 $\rm{mag\ arcsec^{-2}}$. Only 1\% UDGs show S\'ersic indices of $n>2$. 

Sample~2: UDGs selected by \cite{Pina19}, located both the inner (within the virial radii) and outer (beyond the virial radii) regions of eight low-$z$ clusters. UDGs were selected with $r_{\rm{e}}>1.5$~kpc, $\langle\mu_{\rm{e}}(r)\rangle>24\ \rm{mag\ arcsec^{-2}}$, and $n<4$ (only $<3\%$ UDGs have $n>2$); these UDGs have the absolute magnitudes of $-18<M_r<-12.5$~mag.

Sample~3: UDGs in eight low-$z$ clusters selected by \cite{vanderBurg16}. Only the UDG candidates with circular effective radii of $r_{\rm{e,c}}=r_{\rm{e}}\sqrt{b/a}\in (1.5,7.0)$~kpc ($b/a$ denotes the elongation of a galaxy), $\langle\mu_{\rm{e}}(r)\rangle\in (24.0,26.5)\ \rm{mag\ arcsec^{-2}}$, and $n<2$ are included. All of these UDGs are distributed within the virial radii of clusters.

Sample~4: the publicly available {\footnote{http://vizier.cfa.harvard.edu/viz-bin/VizieR?-source=J/ApJ/844/157}} UDG sample in the two intermediate-$z$ clusters, Abell~2744 and Abell~S1063 \citep{Lee17}. UDGs were selected with $r_{\rm{e,c}}>1.5$~kpc and $\langle\mu_{\rm{e,abs}}(r)\rangle>23.8\ \rm{mag\ arcsec^{-2}}$ \citep[$\langle\mu_{\rm{e,abs}}(r)\rangle=\langle\mu_{e,z}(r)\rangle-10\log(1+z)-E(z)-K(z)$ \citep{Graham05}, where $\langle\mu_{\rm{e,abs}}(r)\rangle$ and $\langle\mu_{\rm{e,z}}(r)\rangle$ are the mean surface brightness at $z=0$ and $z$, respectively, and the values of $E(z)$ and $K(z)$ are -0.36 and +0.11 for Abell~S1063, and -0.32 and +0.09 for Abell~2744, respectively; cf.][]{Lee17} which corresponds to the surface brightness criterion of sample~3, i.e., $\langle\mu_{\rm{e,z=0.055}}(r)\rangle>24.0\ \rm{mag\ arcsec^{-2}}$. 90\% UDGs show $n<2$.

The apparent axis ratio $q=b/a$ and its error for the spheroid of each UDG in the four studies were obtained from {\textsc{GALFIT}} \citep{Peng02,Peng10} fitting with a S\'ersic profile{\footnote{These UDGs have been visually inspected to verify whether {\textsc{GALFIT}} provides a proper fit to the data; the bad ones were abandoned by those authors.}}, by the authors of the corresponding studies. 

\subsection{Apparent axis ratios of UDGs}\label{sec:q_dis}


\begin{table*}[!] \footnotesize
	\centering

\begin{tabular}{|c|c|c|c|}
\hline
\hline
\multicolumn{3}{|c|}{Compared subsample pairs in K-S tests} & K-S $p$ value \\
 \hline
\multirow{9}{8em}{low-$z$} & \multirow{3}{10em}{$M_r<-15.2$~mag} & $R\leq 0.5R_{200}$ \textbf{vs.} $0.5R_{200}<R\leq R_{200}$ & $8.2\times10^{-1}$ \\
 &  & $R\leq 0.5R_{200}$ \textbf{vs.} $R>R_{200}$ & $1.2\times10^{-2}$\\
 &  & $0.5R_{200}<R\leq R_{200}$ \textbf{vs.} $R>R_{200}$ & $3.6\times10^{-2}$ \\
 \cline{2-4}
 & \multirow{3}{10em}{$M_r>-15.2$~mag} & $R\leq 0.5R_{200}$ \textbf{vs.} $0.5R_{200}<R\leq R_{200}$ & $3.7\times10^{-1}$ \\
 &  & $R\leq 0.5R_{200}$ \textbf{vs.} $R>R_{200}$ & $1.6\times10^{-4}$\\
 &  & $0.5R_{200}<R\leq R_{200}$ \textbf{vs.} $R>R_{200}$ & $7.2\times10^{-3}$\\
 \cline{2-4}
 & \multirow{1}{10em}{$R\leq 0.5R_{200}$} & $M_r<-15.2$~mag \textbf{vs.} $M_r>-15.2$~mag & $6.8\times10^{-2}$ \\
 \cline{2-4}
 & \multirow{1}{10em}{$0.5R_{200}<R\leq R_{200}$} & $M_r<-15.2$~mag \textbf{vs.} $M_r>-15.2$~mag & $6.9\times10^{-4}$ \\
 \cline{2-4}
 & \multirow{1}{10em}{$R>R_{200}$} & $M_r<-15.2$~mag \textbf{vs.} $M_r>-15.2$~mag & $1.7\times10^{-3}$ \\
\hline
\multirow{2}{8em}{$R\leq R_{200}$} & \multirow{1}{10em}{$M_r<-15.2$~mag} & low-$z$ \textbf{vs.} intermediate-$z$ & $2.6\times10^{-1}$\\
\cline{2-4}
 & \multirow{1}{10em}{$M_r>-15.2$~mag} & low-$z$ \textbf{vs.} intermediate-$z$ & $1.8\times10^{-1}$\\
\hline
\hline
\end{tabular}
\caption{$p$ values from two-sample K-S tests for the different UDG samples. In each line, we compare the axis ratios of the two UDG subsamples selected with the criteria listed in the first three columns. The first two columns show the common properties of the two compared subsamples, while the third column exhibits the different properties of the two compared subsamples. For example, the first line compares the (low-$z$ \& $M_r<-15.2$~mag \& $R\leq 0.5R_{200}$) subsample and (low-$z$ \& $M_r<-15.2$~mag \& $0.5R_{200}<R\leq R_{200}$) subsample, deriving the K-S test $p=8.2\times10^{-1}$; the last line compares the ($R\leq R_{200}$ \& $M_r>-15.2$~mag \& low-$z$) subsample and ($R\leq R_{200}$ \& $M_r>-15.2$~mag \& intermediate-$z$) subsample, deriving the K-S $p=1.8\times10^{-1}$.}
\label{KS}
\end{table*}

\begin{figure*}[!]
\centering
\includegraphics[width=\textwidth]{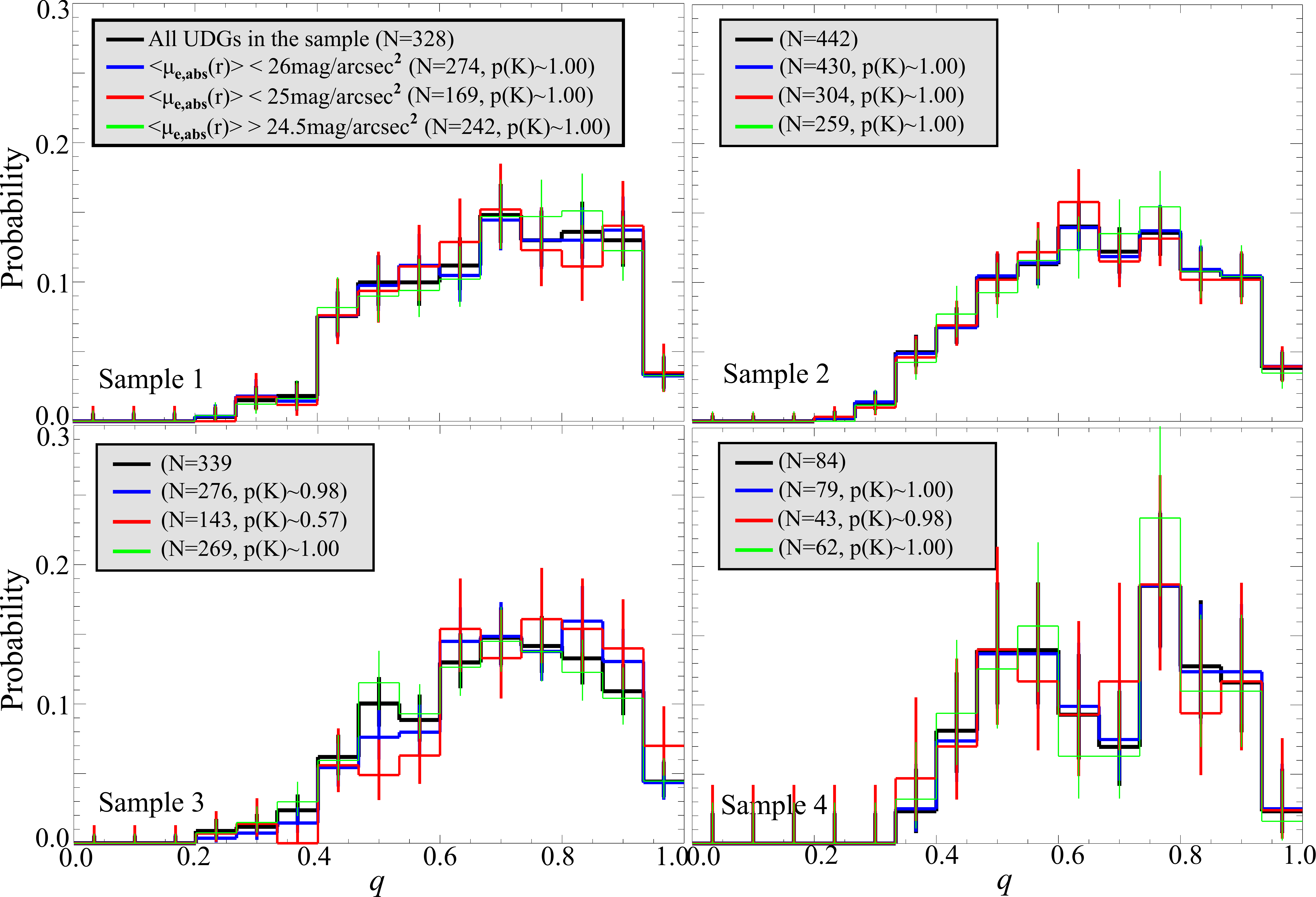}
\caption{The four panels show the distributions of $q$ for the four different UDG samples, described in section~\ref{sec:2.1}. In each panel, the black, blue, red, and green histograms highlight the distributions of $q$ for the whole sample and UDGs with $\langle\mu_{\rm{e,abs}}(r)\rangle<26.0\ \rm{mag\ arcsec^{-2}}$, $\langle\mu_{\rm{e,abs}}(r)\rangle<25.0\ \rm{mag\ arcsec^{-2}}$, and $\langle\mu_{\rm{e,abs}}(r)\rangle>24.5\ \rm{mag\ arcsec^{-2}}$, respectively. The values $N$ and $p$(K) in the legends indicate the numbers of UDGs and $p$ values returned from the two-sample Kuiper tests between the entire sample (black) and subsamples (colored) with the different criteria, respectively. Hereafter, the error bars in distributions show the $68\%$ Wilson intervals by assuming the binomial statistics.}
\label{test_completeness}
\end{figure*}

\begin{figure}[!]
\centering
\includegraphics[width=\columnwidth]{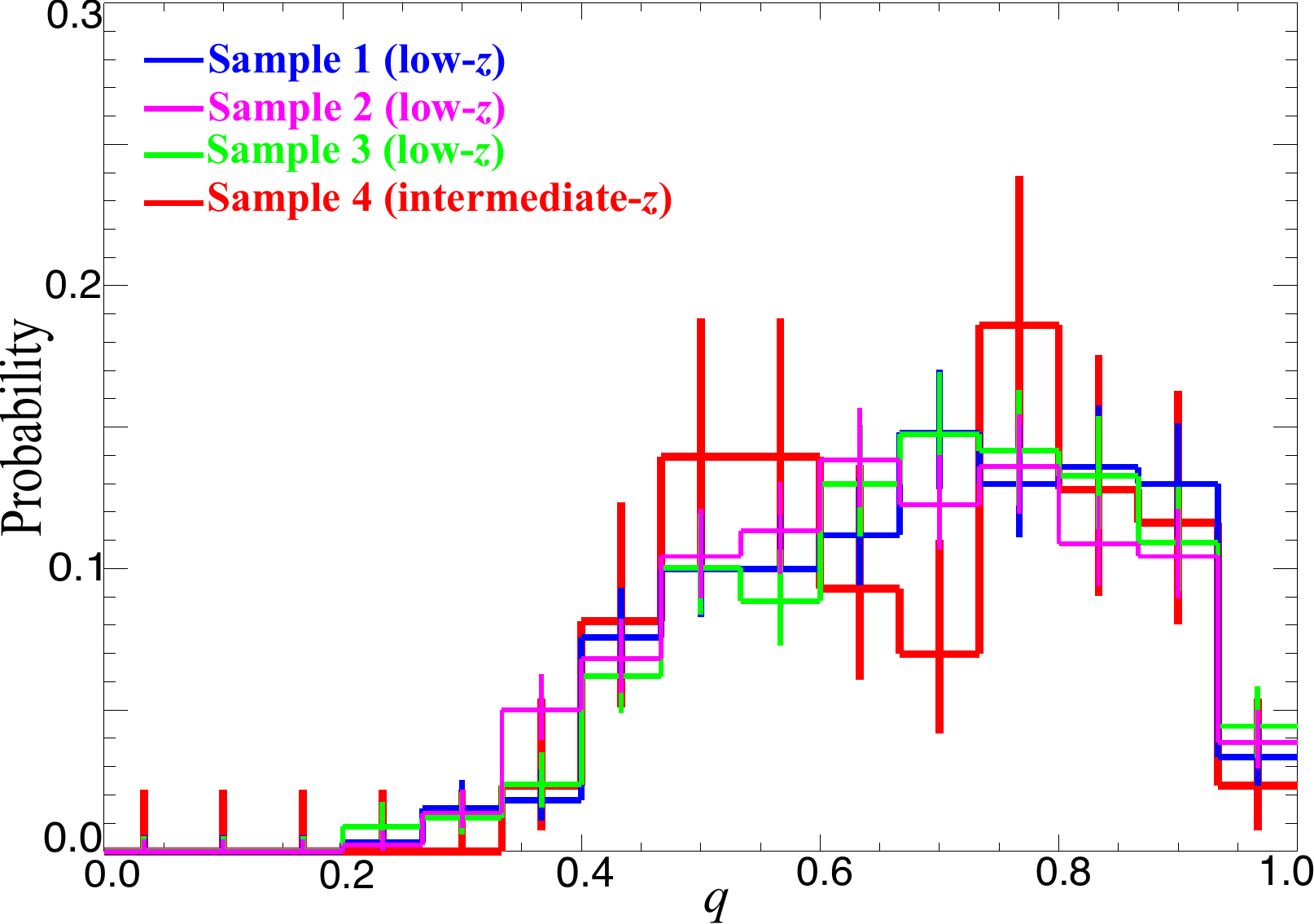}
\caption{The apparent axis ratio $q=b/a$ distributions of the four UDG samples studied in this work.}
\label{q_dis}
\end{figure}

\begin{figure}[!]
\centering
\includegraphics[scale=0.18]{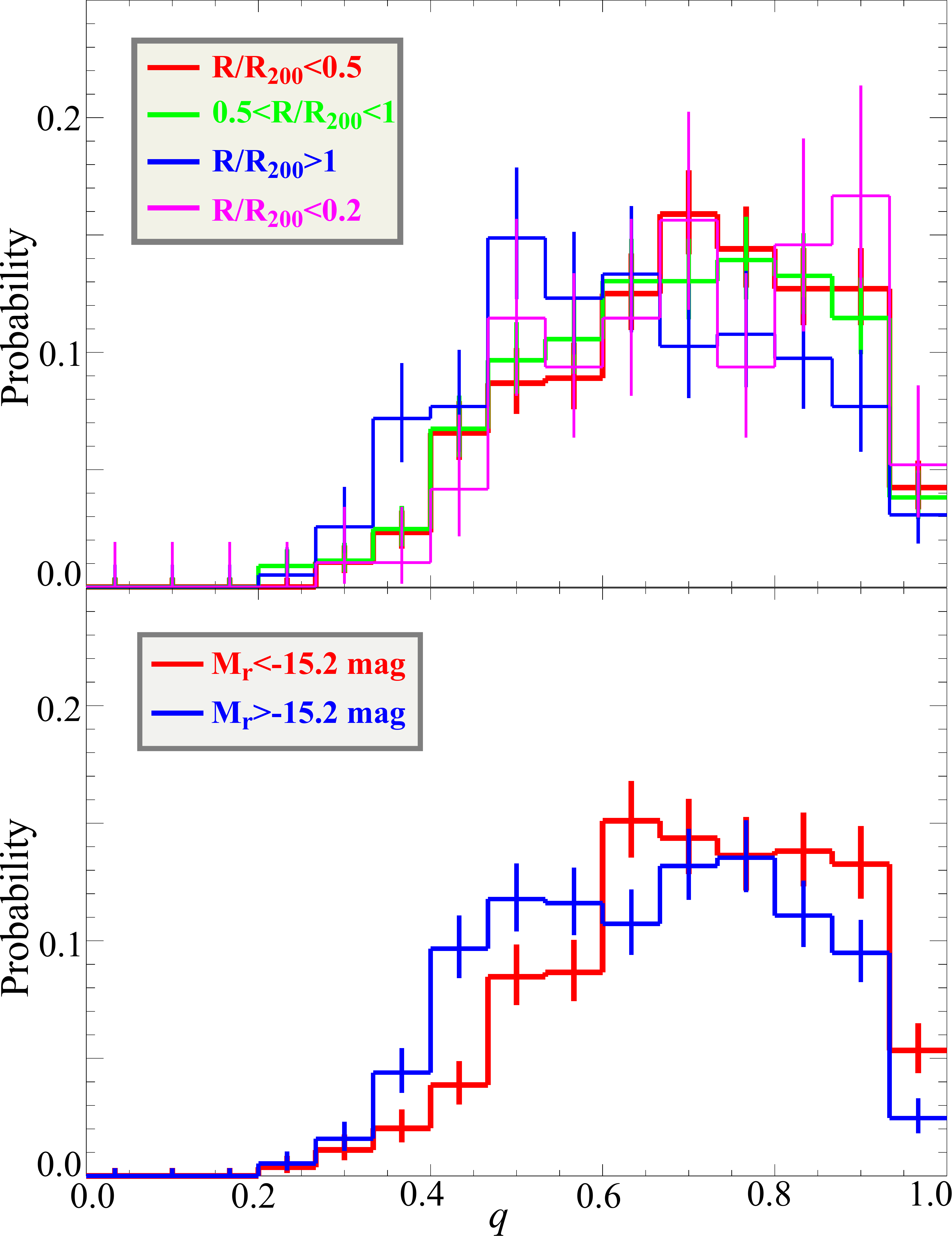}
\caption{Upper: $q$ distributions for the low-$z$ UDGs located in the inner ($R\leq 0.5R_{200}$; red), middle ($0.5R_{200}\leq R<R_{200}$; green), outer ($R>R_{200}$; blue), and innermost ($R<0.2R_{200}$) regions, respectively. Lower: $q$ distributions for the low-$z$ more-luminous ($M_r<-15.2$~mag; red) and less-luminous ($M_r>-15.2$~mag; blue) UDGs, respectively.}
\label{q_dis_divide}
\end{figure}

\begin{figure}[!]
\centering
\includegraphics[scale=0.13]{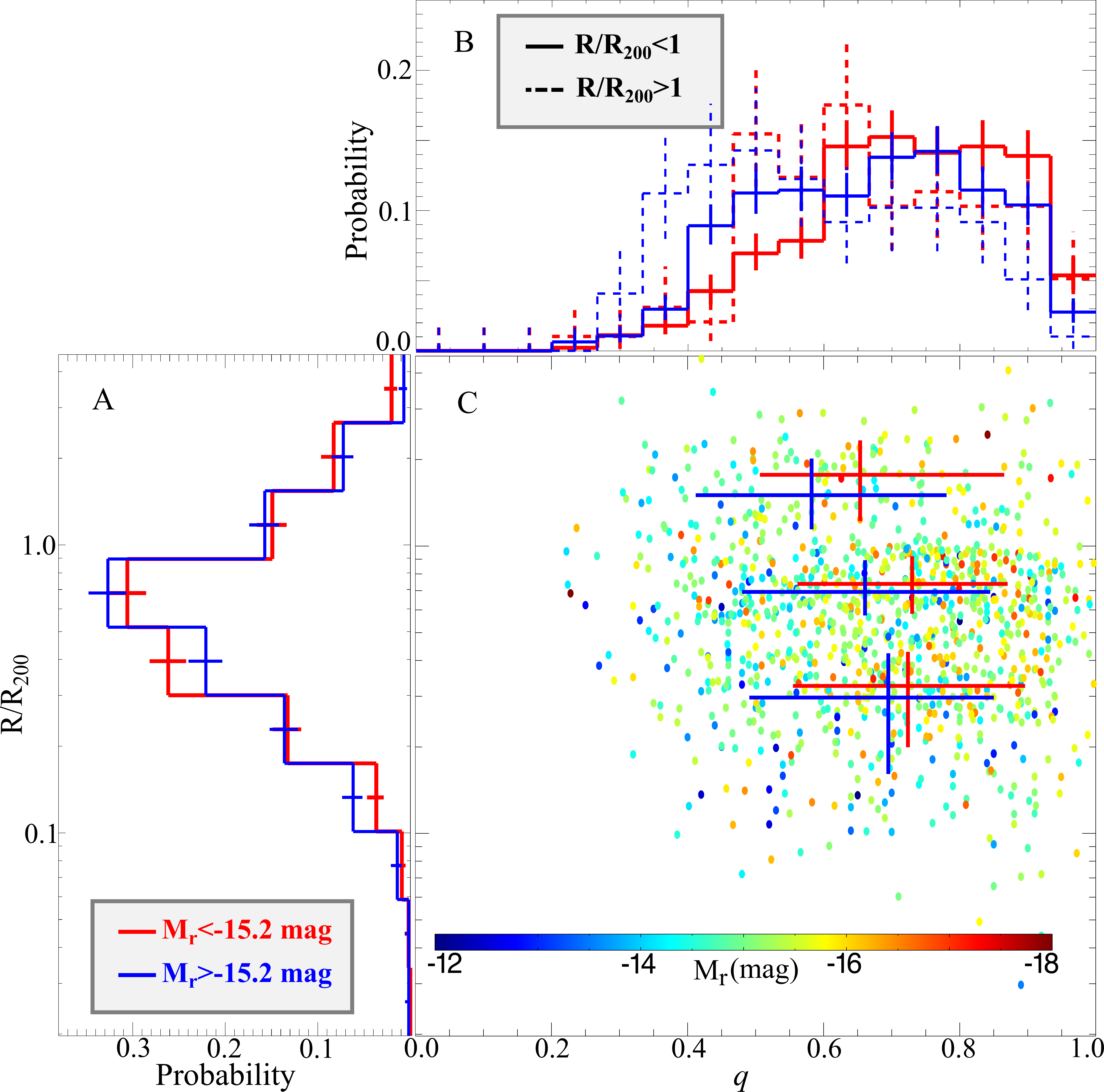}
\caption{Panel~A: the distributions of $R/R_{200}$ for the low-$z$ high-mass ($M_r<-15.2$~mag; red) and low-mass ($M_r>-15.2$~mag; blue) UDGs. Panel~B: the distributions of $q$ for the low-$z$ high-mass (red) and low-mass (blue) UDGs located in $R<R_{200}$ (solid) and $R>R_{200}$ (dashed), respectively. Panel~C: the colored dots show $q$ versus $R/R_{200}$ for the entire low-$z$ UDG sample; their colors denote $M_r$, as shown by the inset color bar; the error bars indicate the median values and $1\sigma$ scatters of $q$ and $R/R_{200}$ for the low-$z$ high-mass (red) and low-mass (blue) UDGs located in $R\leq 0.5R_{200}$, $0.5R_{200}<R\leq R_{200}$, and $R>R_{200}$.}
\label{q_dis_test}
\end{figure}

Note that the four UDG samples were detected by the different telescopes, and thus the faint-ends (for the four samples, the UDG faint-ends of $\langle\mu_{\rm{e,abs}}(r)\rangle$ are $\sim27.0$, 26.5, 26.5, and 26.6~$\rm{mag\ arcsec^{-2}}$, respectively) and detection completeness of the four UDG samples are slightly different. If we assume that the detection completeness of the faintest UDGs is related to their apparent axis ratios (i.e., the edge-on (face-on) oblate (prolate) galaxies perhaps are easier to be detected due to their brighter surface brightness, compared with the face-on (edge-on) ones with the same intrinsic 3D light distribution), the incompleteness of the faint-end UDGs may therefore introduce a bias in the following studies of intrinsic morphologies. Analogously, the four UDG samples adopt a similar UDG bright-end definition of $\langle\mu_{\rm{e,abs}}(r)\rangle >24\ \rm{mag\ arcsec^{-2}}$, which may also cause an absence of the edge-on (face-on) oblate (prolate) UDGs and lead to a bias in this work. Therefore, for each UDG sample, we change the surface brightness faint-end and bright-end of selecting UDGs and test whether the distribution of the apparent axis ratios $q=b/a$ changes with the different criteria. As explored in Fig.~\ref{test_completeness}, for each UDG sample, the $q$ distributions (the $68\%$ Wilson interval \citep{Wilson27,Brown01} for the probability of each bin is estimated, as shown by the error bar in the distribution) for the different faint-ends and bright-ends are consistent with each other within the $1\sigma$ uncertainties; the two-sample Kuiper tests between the subsamples (colored) with the different criteria and entire sample (black) also return large $p$ values, as shown in Fig.~\ref{test_completeness}, suggesting that the faint- and bright-ends for the four UDG samples will not affect our following studies of UDG intrinsic morphologies.

In Fig.~\ref{q_dis}, we compare the $q$ distributions of UDGs in the four different samples. We find that the three low-$z$ UDG samples show very similar $q$ distributions, i.e., flat in the range of $q\in[0.4,0.9]$ but decrease drastically in the ranges of $q<0.4$ and $q>0.95$. Therefore, hereafter the three samples will be combined and treated as one low-$z$ UDG sample in the following studies. We also find that, compared with the $q$ distributions of the low-$z$ samples (the blue, magenta, and green histograms), the intermediate-$z$ UDG sample (the red histogram) exhibits a plausibly less-flat distribution, which resembles a `double-peak' distribution peaking at $q\simeq 0.53$ and 0.77, respectively. It may imply a UDG morphology evolution with redshifts.

It has been known that the properties of UDGs are related to environment and stellar mass \citep{Roman17a,Gu18}. For the low-$z$ UDGs, in order to investigate whether their morphologies evolve from the outside to the inner regions of galaxy clusters, we split the low-$z$ UDGs into three groups, i.e., the inner sample within $R/R_{200}\leq 0.5$ ($R$ denotes the projected distance from a galaxy to the center of its parent cluster), middle sample within $0.5<R/R_{200}\leq 1.0$, as well as outer sample within $R/R_{200}>1.0$. Since the lack of UDGs in the innermost cluster regions has been reported \citep[e.g.,][]{vanderBurg16,Rong17,Pina18} and we may expect a distinct UDG axis ratio distribution, we also plot the $q$ distribution for the innermost sample within $R<0.2R_{200}$. As shown in the upper panel of Fig.~\ref{q_dis_divide}, from the outer to inner region, the median value of the apparent axis ratios increases with the decreasing $R/R_{200}$, from $q_{\rm{median}}=0.61^{+0.22}_{-0.17}$ for the outer sample, to $0.69^{+0.16}_{-0.21}$ for the middle sample, and finally to $0.70^{+0.17}_{-0.19}$ for the inner sample (for the innermost sample, $q_{\rm{median}}=0.70^{+0.19}_{-0.17}$, similar to that of the inner-region sample), suggesting that UDGs become rounder towards the denser environments, which is possibly caused by tidal stripping and heating \citep{Moore96,Aguerri09,Rodriguez-Gomez17,Lisker06,Jiang18,Carleton19,Ogiya18}, or stellar feedback \citep{Teyssier13,Pontzen12,El-Badry16}, etc. For the intermediate-$z$ UDG sample, because of the small galaxy number, we cannot split them into the different $R/R_{200}$ ranges and study their possible morphology evolution with the weak statistical power.

We also divide the low-$z$ UDG sample into the two subsamples with the relatively high ($M_r<-15.2$~mag) and low ($M_r>-15.2$~mag) luminosities (the low-$z$ UDG luminosities range in $M_r\in (-18,-12)$~mag, and $M_r=-15.2$~mag is the median luminosity; cf. Fig.~\ref{lowz_intz_mag}), and compare their $q$ distributions in the lower panel of Fig.~\ref{q_dis_divide}. UDGs with the lower luminosities seem to be more elliptical compared with the ones with the relatively higher luminosities. Since the low-$z$ cluster UDGs show similar colors \citep[e.g.,][]{Rong17}, we can roughly assume a uniform stellar mass-to-light ratio{\footnote{$M_*/L_r\sim 1.96$ \citep{Carleton19}; $M_r\in (-18,-12)$~mag roughly corresponding to $(10^7,10^9)\ M_{\odot}$; $M_r=-15.2$~mag roughly corresponding to $10^{8.2}\ M_{\odot}$.}} for these UDGs; therefore, the results also suggest that the high-mass UDGs are rounder than the low-mass ones.

Note that, if the spatial distribution preferences of the high-mass and low-mass UDGs are different, the dependence of $q$ on luminosity/stellar mass may be actually caused by the dependence on environment, or vice versa. 
Therefore, we compare the distributions of $R/R_{200}$ for the high-mass and low-mass UDG samples, as shown in panel~A of Fig.~\ref{q_dis_test}; we also divide both of the high-mass and low-mass UDG samples into the inner ($R<R_{200}$) and outer ($R>R_{200}$) subsamples, and compare their axis ratios in panels~B and C of Fig.~\ref{q_dis_test}. Apparently, the spatial distributions of the high-mass and low-mass UDGs are barely different (panel~A), whereas the $q$ distributions for the four subsamples with the different luminosities and locations are quite different (panel~B). We find that, both the high-mass and low-mass UDG samples located in $R>R_{200}$ are more elliptical compared with their counterparts located in $R<R_{200}$; both of the inner-region and outer-region low-mass UDGs are more elliptical compared with the high-mass counterparts. The results indicate that the morphologies of low-$z$ UDGs depend on both of the luminosity and environment.

We also use two-sample Kolmogorov-Smirnov (K-S) tests to evaluate the differences between the $q$ distributions of the UDG subsamples within the different luminosity ranges, $R/R_{200}$ ranges, as well as redshifts. The $p$ values from the K-S tests are listed in Table~\ref{KS}. Using the Bonferroni correction with an overall significance level of $95\%$ \citep{Miller66}, we find that, for the low-$z$ UDGs, there may be relatively-significant $q$ differences between the low-mass UDG subsamples located in $R\leq 0.5R_{200}$ and $R>R_{200}$, and between the $R\geq0.5R_{200}$ UDG subsamples with the relatively low and high masses; in addition, the $q$ differences between the high-mass UDG subsamples located in and beyond $R_{200}$ may be moderate (considering the Bonferroni correction with an overall significance level of $68\%$). Yet, the low-$z$ UDGs in $R<0.5R_{200}$ and $0.5R_{200}<R<R_{200}$ show weak/no $q$ differences; the differences between the low-$z$ and intermediate-$z$ UDG counterparts are also mild. In general, the K-S test results imply that the morphologies of UDGs may be related to their luminosities and environments (i.e., located in or beyond $R_{200}$); whereas the morphologic dependence on redshift may be mild/negligible.

\section{Intrinsic morphologies of UDGs}\label{sec:3}

\subsection{Modelling}

\begin{figure*}[!]
\centering
\includegraphics[width=\textwidth]{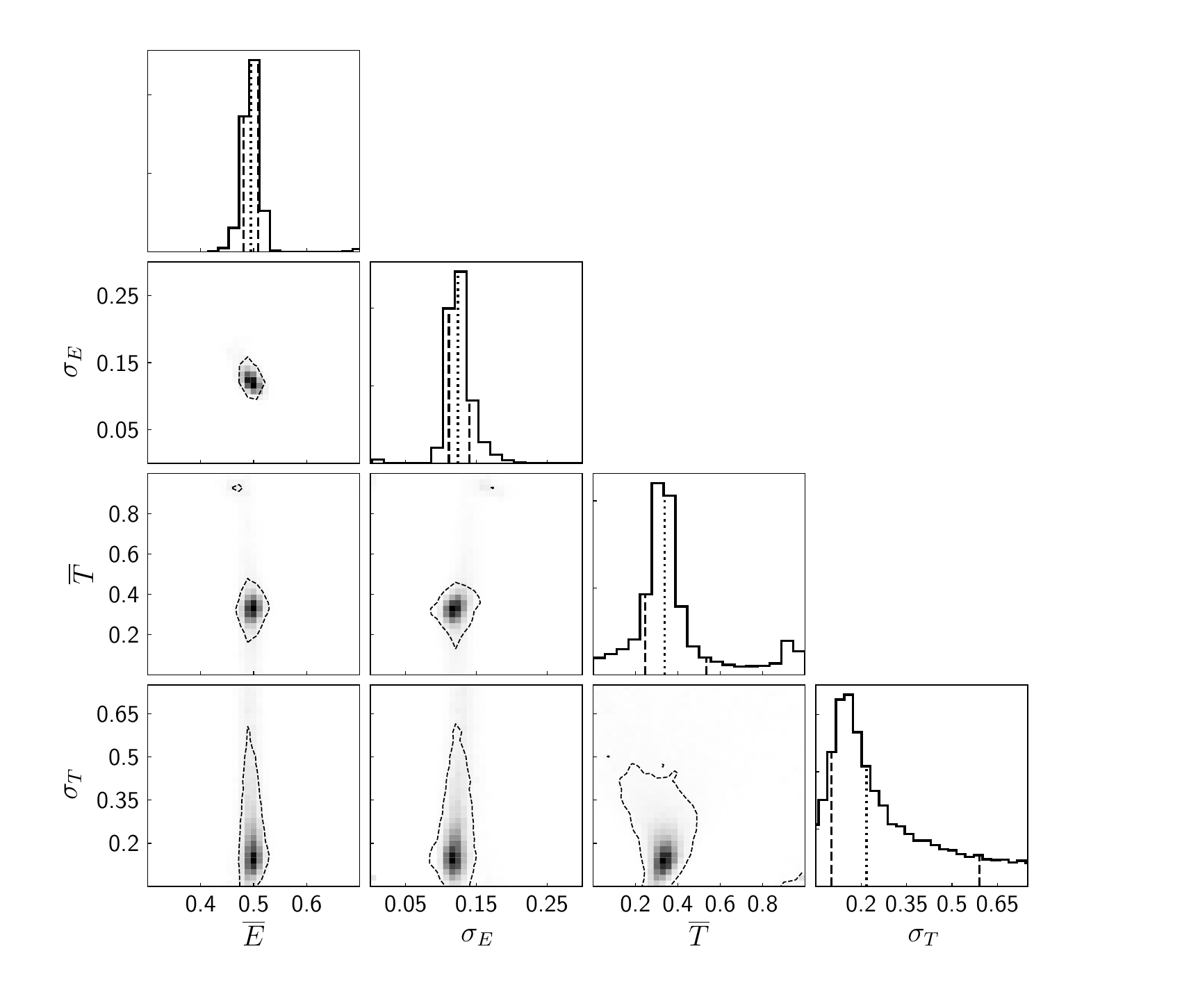}
\caption{Posterior probability density functions of $\bar{E}$, $\sigma_E$, $\bar{T}$, and $\sigma_T$ for sample~1 of low-$z$ UDGs in the Coma cluster. The panels in the diagonal show the posterior pdfs for each of the parameters, marginalized over all the other ones. The grey scale in the non-diagonal panels shows the corresponding joint posterior pdfs. Contours enclose the regions that contain 68\% of the cumulative posterior probability. The dotted and dashed lines in the diagonal panels indicate the 50\% and 16\% and 84\% of the corresponding marginalized posteriors, respectively.}
\label{MC_Yagi}
\end{figure*}

\begin{figure*}[!]
\centering
\includegraphics[width=\textwidth]{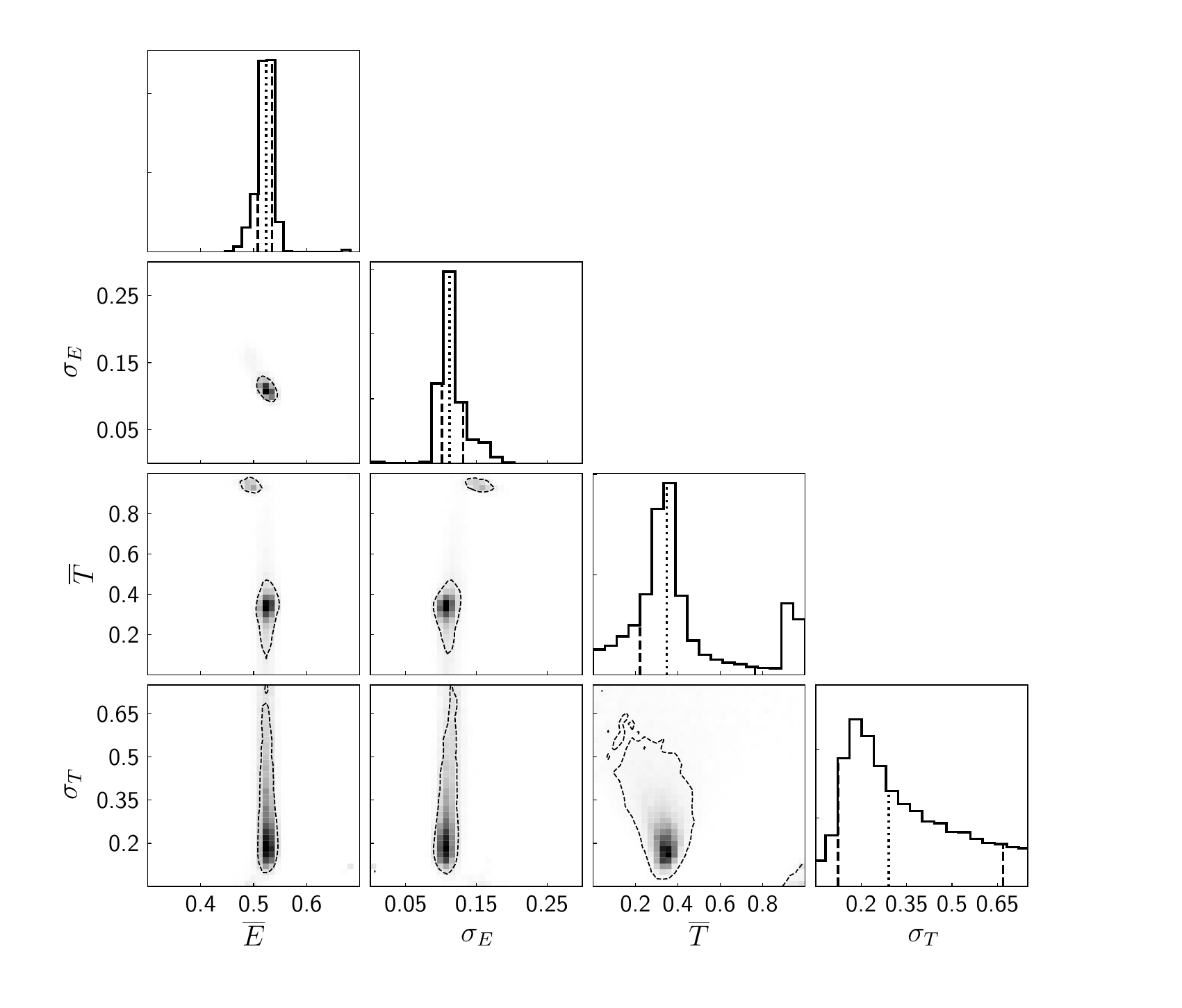}
\caption{Analogous to Fig.~\ref{MC_Yagi}, posterior probability density functions of $\bar{E}$, $\sigma_E$, $\bar{T}$, and $\sigma_T$ for sample~2 of low-$z$ UDGs in 8 nearby clusters.}
\label{MC_Pina}
\end{figure*}

\begin{figure*}[!]
\centering
\includegraphics[width=\textwidth]{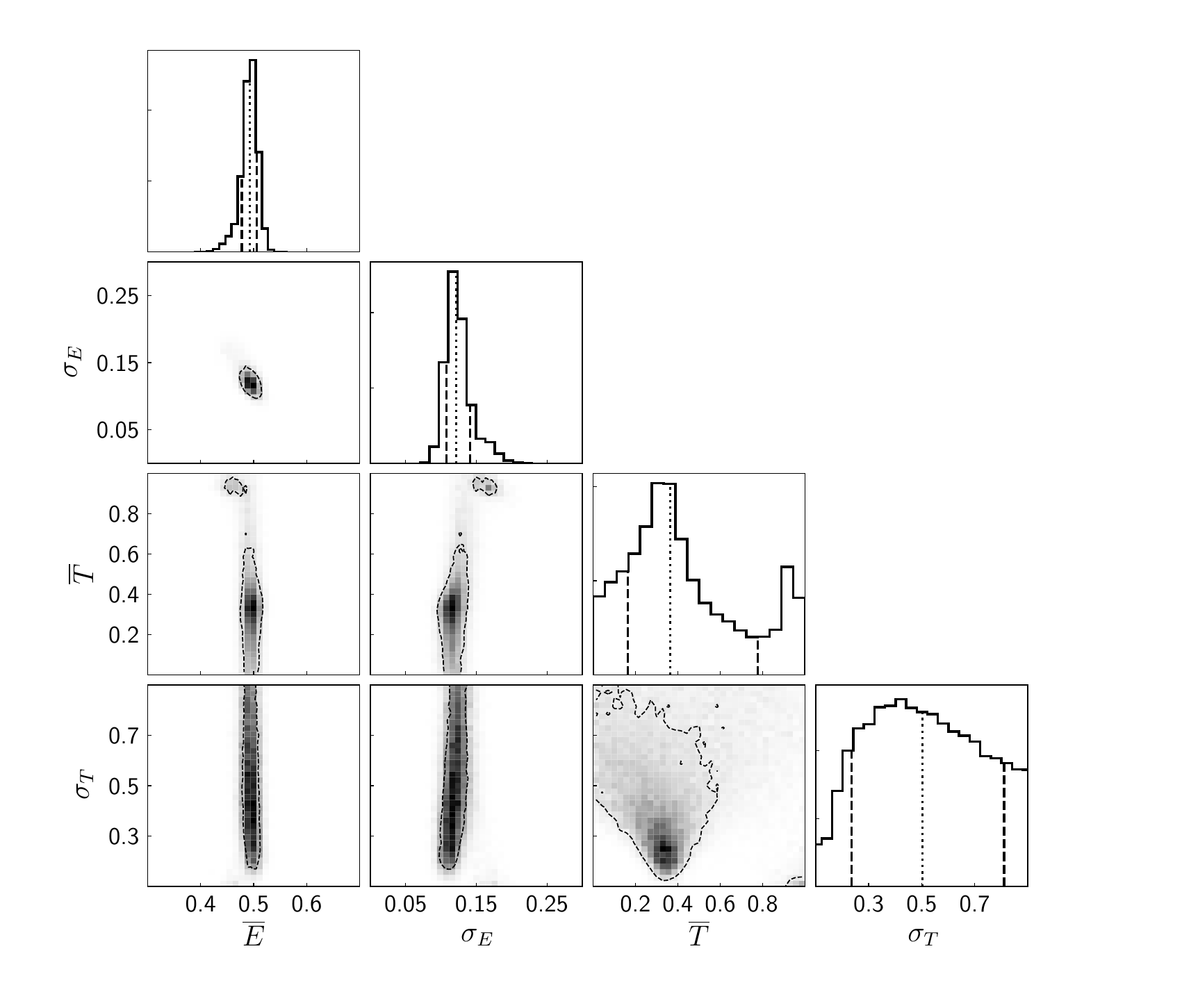}
\caption{Analogous to Fig.~\ref{MC_Yagi}, posterior probability density functions of $\bar{E}$, $\sigma_E$, $\bar{T}$, and $\sigma_T$ for sample~3 of low-$z$ UDGs in 8 nearby clusters.}
\label{MC_Burg8}
\end{figure*}

\begin{figure*}[!]
\centering
\includegraphics[width=\textwidth]{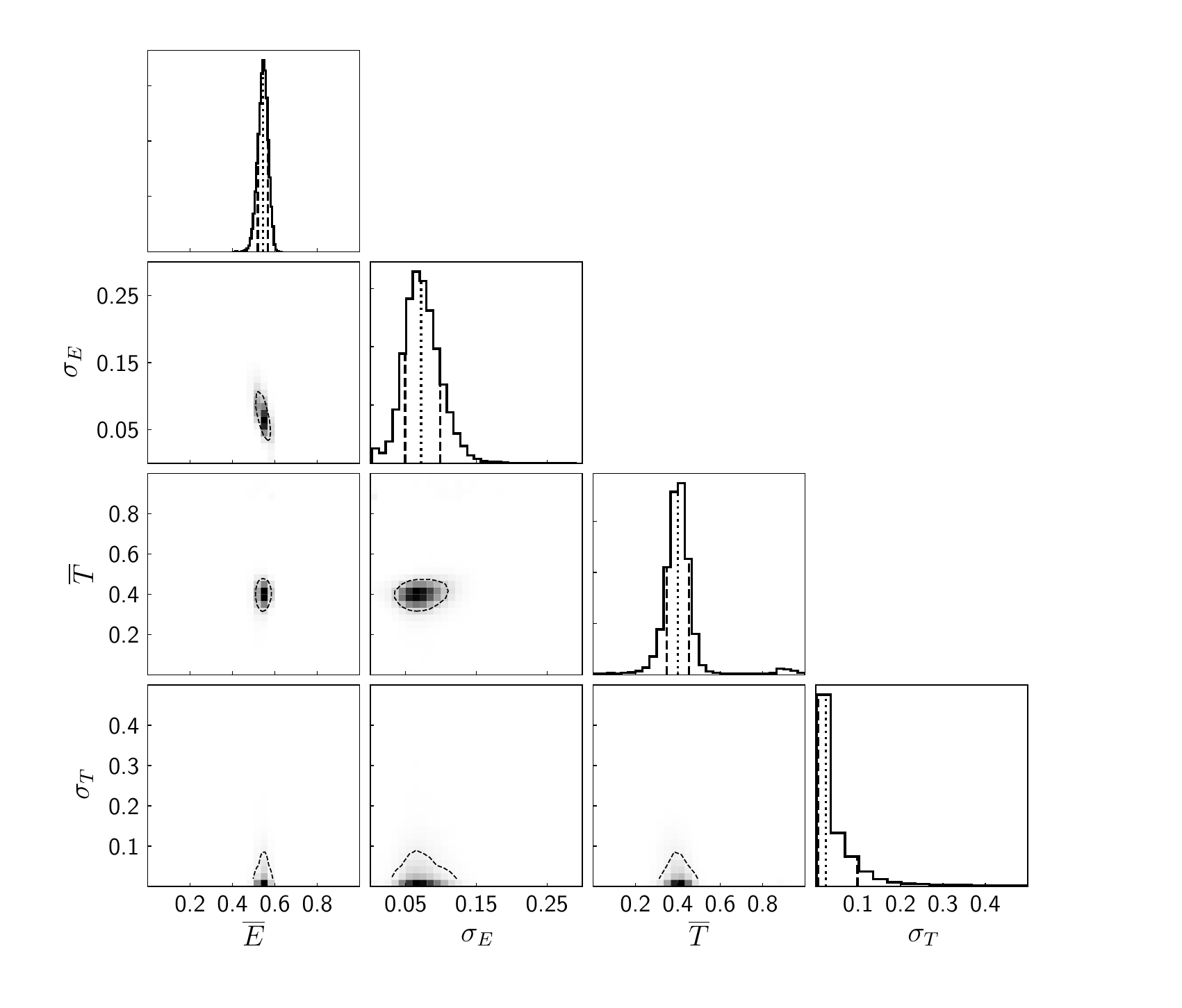}
\caption{Analogous to Fig.~\ref{MC_Yagi}, posterior probability density functions of $\bar{E}$, $\sigma_E$, $\bar{T}$, and $\sigma_T$ for sample~4 of intermediate-$z$ UDGs in Abell~2744 and Abell~S1063.}
\label{MC_interz}
\end{figure*}

In this section, we will take advantage of a Markov Chain Monte Carlo (MCMC) approach to investigate the intrinsic morphologies of UDGs in the different samples. The method described in \cite{Sanchez-Janssen16} is used to analyze the intrinsic morphologies of UDGs. Here we briefly outline the key points of the method. In this method, the galaxies in each sample are modelled as a family of optically-thin triaxial ellipsoids. The 3D galaxy density is structured as a set of coaligned ellipsoids characterized by a common ellipticity $E=1-C/A$, and a triaxiality $T = (A^2-B^2)/(A^2-C^2)$, where $A\geq B\geq C$ are the intrinsic major, intermediate, and minor axes of the ellipsoid, respectively \citep{Franx91}. The purely prolate (oblate) model corresponds to $T\simeq1$ ($T\simeq0$).

For each UDG sample, their $E$ and $T$ are proposed to follow Gaussian distributions, with mean values and standard deviations of $\bar{E}$, $\sigma_E$, $\bar{T}$, and $\sigma_T$. Given the distribution of intrinsic axis ratios and random viewing angles for the model galaxies, the distribution of apparent axis ratios $q$ can be derived via projecting these ellipsoids \citep[cf. the Appendix for the projecting method details]{Rong15a}. Therefore conversely, the posterior probability density function (pdf) of the model parameters $\bar{E}$, $\sigma_E$, $\bar{T}$, and $\sigma_T$, representing the intrinsic shapes of each UDG population can also be inferred by applying a Bayesian framework \citep[cf.][]{Sanchez-Janssen16}, and assuming the prior probabilities of $\bar{E}$ and $\bar{T}$ to follow uniform distributions in $[0,1]$, as well as $\sigma_E$ and $\sigma_T$ to follow $p(\sigma)\propto \sigma^{-1}$. We implement the {\textsc{emcee}} code \citep{Foreman-Mackey13} to sample the posterior distribution of the model parameters with 200 `walkers' and 1500 steps (the steps are sufficient for the MCMC chains to reach equilibrium).

The modelling results for the different UDG samples are summarized in Table~\ref{UDG_morphology}. In Figs.~\ref{MC_Yagi}, \ref{MC_Pina}, \ref{MC_Burg8}, and \ref{MC_interz}, we plot the posterior distributions of $\bar{E}$, $\sigma_E$, $\bar{T}$, and $\sigma_T$ for the low-$z$ and intermediate-$z$ UDG samples, respectively. The posterior distributions for $\bar{E}$ and $\sigma_E$ approximately resemble the single Gaussian distributions. However, for the low-$z$ UDG samples, the posterior distributions of $\bar{T}$ always show double-peaks (cf. also Appendix~\ref{sec:aa}), i.e., one pronounced peak at $\bar{T}\sim 0.3\--0.4$ and one weak peak at $\bar{T}\sim 1.0$; for the intermediate-$z$ sample, the posterior distribution of $\bar{T}$ exhibits no significant $\bar{T}\sim 1.0$ peak. The two $\bar{T}$ peaks indicate two different probable underlying shapes for the low-$z$ UDGs: the left $\bar{T}\sim 0.3\--0.4$ peak corresponds to a triaxial model, and right $\bar{T}\sim 1.0$ peak corresponds to a purely-prolate model. We note that, although the precise triaxiality distribution can only be well derived with a synergy of structural and kinematical data \citep[e.g.,][]{Franx91,Bosch09,Rong18,Chilingarian19}, rather than with the photometric data alone, yet the data clearly favor triaxial models over nearly prolate ones.{\footnote{We also note that, in theory, one can alternatively assume that one UDG sample is made up of two divergent populations with $\bar{T}\sim 0.3\--0.4$ and $\bar{T}\sim 1.0$, respectively; however, we prefer a simpler model with only one set of ellipticity and triaxiality parameters to describe the intrinsic morphology of cluster UDGs in one sample, particularly in the case that the triaxial model alone can well recover the $q$ distributions of UDGs. }} The discovery of the triaxial UDG morphologies is in consistent with the hydrodynamical simulation results \citep[e.g.,][]{Jiang18} but in conflict with the conclusion of \cite{Burkert17}. This conflict is due to the fact that \cite{Burkert17} only compared the purely-oblate and purely-prolate models, while in nature there is no a-priori reason for preferring either of these extremes.

\begin{table*}[!] 
\begin{tabular}{@{}c|ccccccc@{}}
\hline
\hline
Redshift & UDG samples & $N$ & $\bar{E}$ & $\sigma_E$ & $\bar{T}$ & $\sigma_T$ & $A:B:C$ \\
\hline
 & All (Sample 1+2+3) & 1109 & $0.51_{-0.02}^{+0.01}$ & $0.12_{-0.01}^{+0.02}$ & $0.34_{-0.05}^{+0.22}$ & $0.21_{-0.08}^{+0.15}$ & $1:0.86:0.49$ \\
 & Sample~1 & 328 & $0.49_{-0.01}^{+0.01}$ & $0.12_{-0.01}^{+0.02}$ & $0.34_{-0.09}^{+0.20}$ & $0.22_{-0.12}^{+0.37}$ & $1:0.86:0.51$ \\
 & Sample~2 & 442 & $0.52_{-0.02}^{+0.01}$ & $0.11_{-0.01}^{+0.02}$ & $0.35_{-0.13}^{+0.41}$ & $0.29_{-0.17}^{+0.38}$ & $1:0.85:0.48$ \\
 & Sample~3 & 339 & $0.49_{-0.02}^{+0.01}$ & $0.12_{-0.01}^{+0.02}$ & $0.36_{-0.20}^{+0.41}$ & $0.50_{-0.27}^{+0.31}$ & $1:0.86:0.51$ \\
 & $R\leq 0.5R_{200}$ & 471 & $0.48_{-0.01}^{+0.01}$ & $0.12_{-0.01}^{+0.01}$ & $0.37_{-0.12}^{+0.29}$ & $0.36_{-0.19}^{+0.37}$ & $1:0.85:0.52$ \\
 & $0.5R_{200}<R\leq R_{200}$ & 443 & $0.50_{-0.02}^{+0.01}$ & $0.12_{-0.01}^{+0.02}$ & $0.35_{-0.15}^{+0.38}$ & $0.38_{-0.20}^{+0.36}$ & $1:0.86:0.50$ \\
 & $R>R_{200}$ & 195 & $0.57_{-0.02}^{+0.02}$ & $0.22_{-0.01}^{+0.02}$ & $0.37_{-0.18}^{+0.55}$ & $0.39_{-0.34}^{+0.38}$ & $1:0.84:0.43$ \\
low-$z$ UDGs & $R\leq 0.2R_{200}$ & 96 & $0.47_{-0.02}^{+0.01}$ & $0.12_{-0.02}^{+0.03}$ & $0.29_{-0.07}^{+0.09}$ & $0.06_{-0.05}^{+0.22}$ & $1:0.89:0.53$ \\
 & $M_r<-15.2$ & 543 & $0.49_{-0.01}^{+0.01}$ & $0.12_{-0.01}^{+0.01}$ & $0.35_{-0.08}^{+0.14}$ & $0.24_{-0.10}^{+0.37}$ & $1:0.86:0.51$ \\
 & $M_r>-15.2$ & 566 & $0.52_{-0.01}^{+0.01}$ & $0.11_{-0.01}^{+0.01}$ & $0.33_{-0.15}^{+0.21}$ & $0.37_{-0.18}^{+0.34}$ & $1:0.86:0.48$ \\
 & $M_r<-15.2$ \& $R\leq 0.5R_{200}$ & 230 & $0.45_{-0.02}^{+0.02}$ & $0.13_{-0.02}^{+0.03}$ & $0.38_{-0.18}^{+0.51}$ & $0.34_{-0.29}^{+0.38}$ & $1:0.86:0.55$ \\
 & $M_r<-15.2$ \& $0.5R_{200}<R\leq R_{200}$ & 216 & $0.45_{-0.02}^{+0.02}$ & $0.14_{-0.02}^{+0.03}$ & $0.67_{-0.29}^{+0.22}$ & $0.34_{-0.26}^{+0.36}$ & $1:0.73:0.55$ \\
 & $M_r<-15.2$ \& $R>R_{200}$ & 97 & $0.51_{-0.03}^{+0.02}$ & $0.12_{-0.02}^{+0.03}$ & $0.36_{-0.16}^{+0.55}$ & $0.17_{-0.15}^{+0.47}$ & $1:0.85:0.49$ \\
 & $M_r>-15.2$ \& $R\leq 0.5R_{200}$ & 241 & $0.51_{-0.02}^{+0.02}$ & $0.11_{-0.01}^{+0.02}$ & $0.44_{-0.10}^{+0.32}$ & $0.28_{-0.15}^{+0.38}$ & $1:0.82:0.49$ \\
 & $M_r>-15.2$ \& $0.5R_{200}<R\leq R_{200}$ & 227 & $0.54_{-0.02}^{+0.02}$ & $0.10_{-0.02}^{+0.02}$ & $0.27_{-0.14}^{+0.26}$ & $0.26_{-0.20}^{+0.29}$ & $1:0.89:0.46$ \\
 & $M_r>-15.2$ \& $R>R_{200}$ & 98 & $0.65_{-0.02}^{+0.02}$ & $0.04_{-0.02}^{+0.02}$ & $0.56_{-0.21}^{+0.36}$ & $0.34_{-0.26}^{+0.37}$ & $1:0.71:0.35$ \\
\hline
intermediate-$z$ UDGs & All (Sample~4) & 84 & $0.54_{-0.02}^{+0.02}$ & $0.07_{-0.02}^{+0.03}$ & $0.40_{-0.05}^{+0.05}$ & $0.02_{-0.02}^{+0.07}$ & $1:0.83:0.46$ \\
\hline
\hline
\end{tabular}
\caption{The MCMC results of intrinsic morphology analysis for the different samples of low-$z$ and intermediate-$z$ UDGs.
Col. (1): low-$z$ or intermediate-$z$ UDG sample; Col. (2): the UDG sample with the different properties; Col. (3): number of UDGs in each sample used for MCMC; Col. (4)-(7): the mean values and standard deviations of ellipticity and triaxiality distributions $\bar{E}$, $\sigma_E$, $\bar{T}$, and $\sigma_T$; Col. (8)\--(9): the median ratios of three intrinsic axes $A:B:C$.}
\label{UDG_morphology}
\end{table*}

The intrinsic axis ratios, $C/A$ and $B/A$, are then calculated from $\bar{E}$, $\sigma_E$, $\bar{T}$, and $\sigma_T$. Since $E\sim 0.5\--0.6$ and $T\sim 0.3\--0.4$, we find $C/A\sim 0.4\--0.5$ and $B/A\sim 0.8\--0.9$, suggesting that the intrinsic morphologies of cluster UDGs actually are `oblate-triaxial' ($C/B<B/A$).

In order to study the shape dependence on luminosities, environments, as well as redshifts, we also implement the MCMC method to analyze the intrinsic morphologies of the UDG subsamples with the different properties and show the results in Table~\ref{UDG_morphology}. Since compared with the ellipticity distributions, the triaxiality distributions are worse constrained with photometric data alone \citep{Binggeli80}, analogous to the work of \cite{Sanchez-Janssen19}, we will therefore focus our analysis on the comparison of galaxy flattenings ($\bar{E}$) and thicknesses ($C/A$).

\subsection{UDG morphology evolution with masses}\label{sec:mass}

\begin{figure}[!]
\centering
\includegraphics[width=\columnwidth]{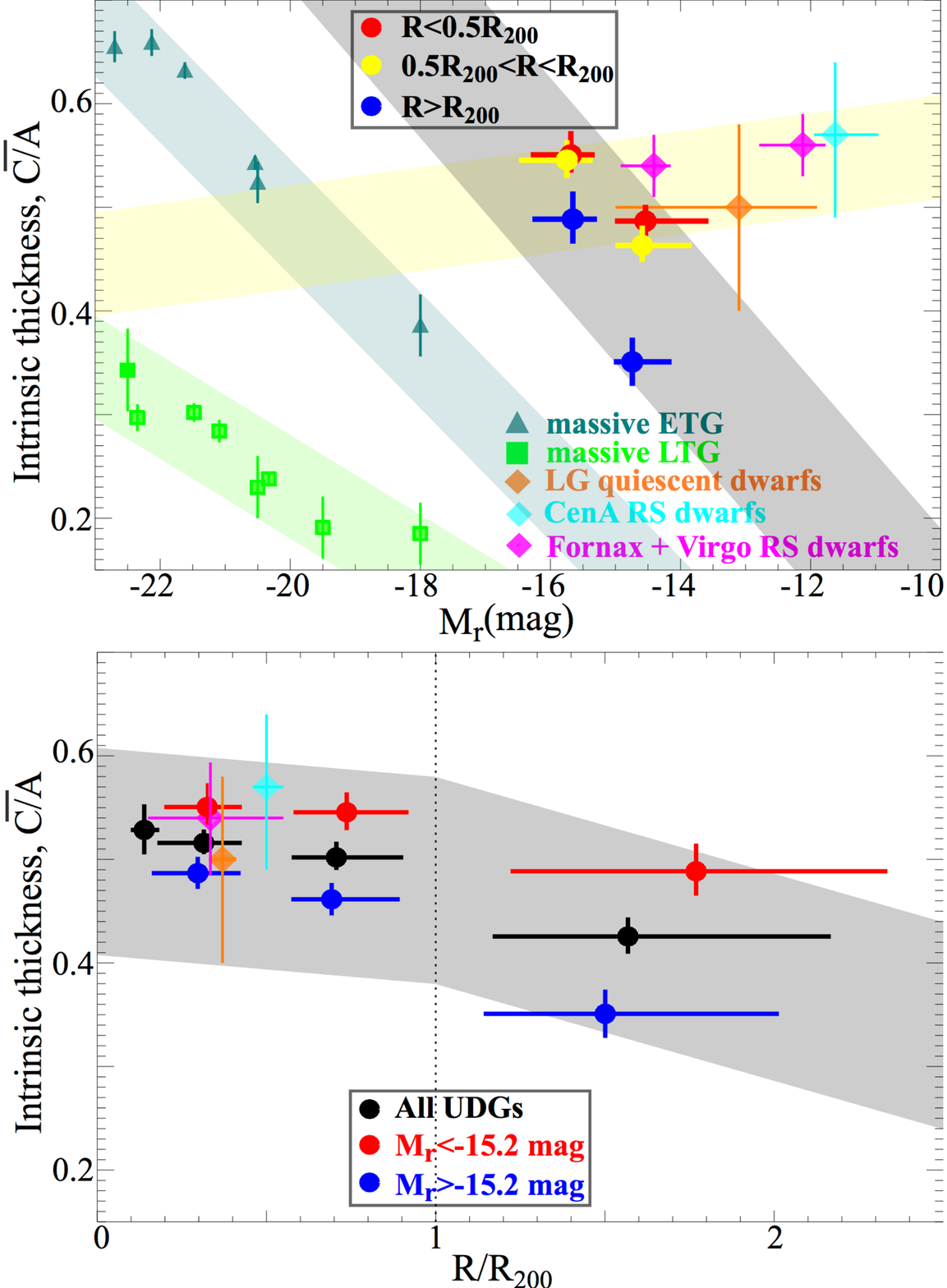}
\caption{The median intrinsic thickness $\bar{C/A}$ as a function of luminosity (upper panel) and $R/R_{200}$ (lower panel) for our low-$z$ UDG samples. In the two panels, the filled circles denote the all, high-mass, and low-mass UDGs, respectively. In the upper panel, the red, yellow, and blue filled circles denote the UDGs located in $R\leq 0.5R_{200}$, $0.5R_{200}<R\leq R_{200}$, and $R>R_{200}$, respectively. In the lower panel, the black, red, and blue filled circles denote the all, high-mass, and low-mass UDGs, respectively, and the vertical dotted line denotes the virial radius of $R=R_{200}$. For comparison, we also plot the thicknesses of quiescent dE/dSphs which follow the red sequence in nearby clusters \citep[Virgo+Fornax; magenta diamond;][]{Sanchez-Janssen19} and groups \citep[Local~Group, orange; Centaurus~A, cyan;][]{McConnachie12,Taylor17,Taylor18,Sanchez-Janssen19}, as well as massive ETGs (dark-green triangles) and LTGs (light-green triangles) selected from SDSS catalog by \cite{Padilla08} and \cite{Rodriguez13}. The gray shaded regions show the thickness trends for the UDG samples with the different luminosities and located in the different environments; the dark- and light-green shaded regions highlight the corresponding thickness trends for the massive ETGs and LTGs within the different luminosity ranges, respectively; the yellow shaded region reveals the thickness trend for the dE/dSphs with the different luminosities in nearby clusters and groups shown by \cite{Sanchez-Janssen19}. The slopes of the plotted gray shaded trends are derived from linear fitting to the point pairs corresponding to the same environments (upper panel) or same mass ranges (lower panel); it is worth to note that the quantitative slopes are not crucial in this work as there are large uncertainties.}
\label{CA_location_mag}
\end{figure}

As listed in Table~\ref{UDG_morphology}, the intrinsic morphologies of the high-mass and low-mass low-$z$ UDGs are significantly different, regardless of their environments. In the upper panel of Fig.~\ref{CA_location_mag}, we compare the median intrinsic thickness, $\bar{C/A}$, for the high-mass and low-mass UDGs located in $R\leq 0.5R_{200}$ (red), $0.5R_{200}<R\leq R_{200}$ (yellow), and $R>R_{200}$ (blue), respectively. We find that, 1) for the UDGs located in the same environment, the high-mass ones are always thicker, puffier, than the low-mass ones (cf. also Appendix~\ref{sec:bb}); 2) the morphology difference between the high-mass and low-mass ones is present in both of $R<R_{200}$ and $R>R_{200}$.

Since the thickness difference between the high-mass and low-mass UDGs always exists in the different environments, it may be originated from internal processes, e.g., supernovae feedback. Star formation in dwarf galaxies is expected to occur in episodic bursts at almost all redshifts \citep{Muratov15}, and the associated supernovae-driven outflows pressurize gas and heat{\footnote{Multiple supernovae explosions induce strong and repeated fluctuations in the dwarf gravitational potential, which result in energy transfer to the collisionless components (dark matter and stars).}} the stellar orbits \citep{Pontzen12,El-Badry16,Teyssier13,Governato10}. If UDGs are proposed to be produced by outflows \citep[e.g.,][]{DiCintio17}, as a consequence of more starbursts or a larger amount of star formation, the higher-mass UDGs should have delivered more energy to heat up the stellar random motions and thus become less flattened.

Galaxy merging can also lead to morphological transformation \citep[e.g.,][]{Starkenburg16a,Starkenburg16b}, and simultaneously, starbursts, which can significantly boost stellar mass assembly. However, many simulations and observational results have excluded mergers to play any significant role in determining the morphologies of dwarf galaxies \citep[e.g.,][]{Rodriguez-Gomez17,Rodriguez-Gomez15,Stewart08}, as the mergers of dwarf galaxies are very rare. 

Although direct mergers are extremely rare, the rate of tidal encounters with more massive galaxies during flybys \citep[including `harassment' and tidal stirring;][]{Moore96,Mayer01,Mayer07} is considerable, particularly in the massive clusters comprising many massive satellites. The tidal interactions can also efficiently puff up UDGs, transform the kinematic and stellar distributions to resemble the present-day dE/dSphs \citep[e.g.,][]{Carleton19,Moore96,Mayer01,Errani15}. In simulations, the stellar dispersions in cluster UDGs with more massive remanent stellar masses after tidal interactions are found to be higher than those in lower-mass counterparts \citep[cf.][]{Carleton19}, plausibly explaining the thicker galactic bodies of the higher-mass UDGs as shown in the upper panel of Fig.~\ref{CA_location_mag}. 

\cite{Sanchez-Janssen19} found that the morphologies of typical quiescent dwarf galaxies, populating the red sequence (RS), also depend on galaxy luminosities. We therefore compare the UDG thickness dependence on luminosity with that of the quiescent dE/dSphs located in the nearby Virgo and Fornax clusters \citep[][]{Sanchez-Janssen19}, and in the Local Group \citep{McConnachie12} and Centaurus~A galaxy group \citep[][]{Taylor17,Taylor18}{\footnote{The $r$-band absolute magnitudes of the dE/dSphs in the Virgo and Fornax clusters as well as Centaurus~A group are obtained by assuming $g-r\sim0.6$ \citep[e.g.,][]{Rong17,Venhola18}, while the magnitudes of the dwarfs in the Local group are directly found in \cite{McConnachie12}.}}, as well as trends of massive early-type galaxies (ETGs) and late-type galaxies (LTGs) selected from the Sloan Digital Sky Survey datasets \citep{Padilla08,Rodriguez13}. As shown in the upper panel of Fig.~\ref{CA_location_mag}, the thickness trend of the low-$z$ cluster UDGs with luminosities (the gray shaded region) is apparently contrary to the trend of the typical quiescent dE/dSphs{\footnote{The thicknesses of the bright and faint dwarfs in \cite{Sanchez-Janssen19} are obtained from $C/A=(C/A_{\rm{nuc}}*N_{\rm{nuc}}+C/A_{\rm{non}}*N_{\rm{non}})/N_{\rm{tot}}$, where $C/A_{\rm{nuc}}, C/A_{\rm{non}}$ denote the thicknesses of the nucleated and non-nucleated dwarfs in each sample, respectively, and $N_{\rm{nuc}}, N_{\rm{non}}, N_{\rm{tot}}$ denote the number of the nucleated, non-nucleated, and entire sample of dwarfs, respectively.}} in nearby clusters and groups \citep[the yellow shaded region; cf. also][]{Sanchez-Janssen19}, but is `akin to' the trends of massive ETGs (the dark-green shaded region) and LTGs{\footnote{There is also literature discussing that the fainter LTGs, with absolute magnitudes of $M_g>-18$~mag, again show thicker morphologies as they get fainter \citep[e.g.,][]{Sanchez-Janssen10,Roychowdhury13}; however in this work, we only focus on the morphology trend for the massive LTGs with $M_g<-18$~mag.}} (the light-green shaded region). Note, the trend (yellow shaded) of typical dE/dSphs is believed to be caused by the lower binding energies of the fainter dwarfs \citep[cf. e.g.,][]{Sanchez-Janssen19}; the different trends of UDGs and typical dE/dSphs may imply that the dramastic internal or external processes, such as stellar feedback or tidal interaction, plays a much more crucial role in the evolution of UDGs, compared with the typical dE/dSphs. In this sense, UDGs may not be simply treated as an extension of the dE/dSph class with similar evolutionary histories; they may differ not only in size.

\subsection{UDG morphology evolution with environments}\label{sec:environment}

\begin{figure}[!]
\centering
\includegraphics[width=\columnwidth]{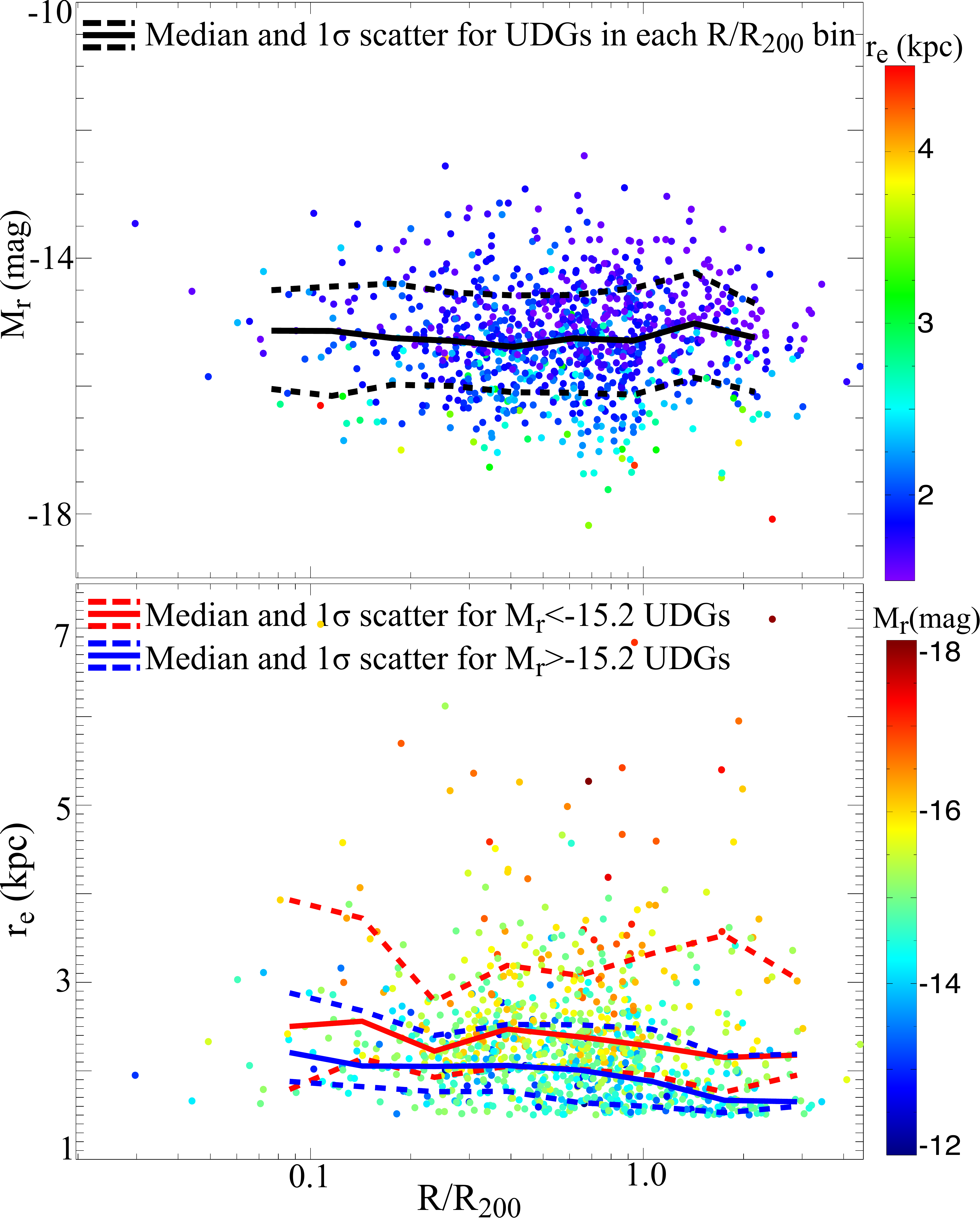}
\caption{The luminosities (upper panel) and UDG sizes (lower panel) as functions of $R/R_{200}$. In the upper panel, the different colored dots denote the low-$z$ UDGs with the different effective radii; the solid and dashed components show the median and $1\sigma$ scatter of UDG luminosities in each $R/R_{200}$ bin. In the lower panel, the different colored dots denote the low-$z$ UDGs with the different $M_r$; the solid and dashed components show the median and $1\sigma$ scatter of UDG sizes in each $R/R_{200}$ bin for the high-mass (red) and low-mass (blue) subsamples, respectively. We only plot the results for the conservative sample of UDGs with $r_{\rm{e,c}}>1.5$~kpc.}
\label{dis_re_mag}
\end{figure}

In the lower panel of Fig.~\ref{CA_location_mag}, we show the median intrinsic thickness as a function of $R/R_{200}$ for the entire (black), high-mass (red), and low-mass (blue) low-$z$ UDG samples. From $R>R_{200}$ to $R<R_{200}$, UDGs become significantly puffed up; this outside-in evolution is stronger for lower-mass systems. However, in the virial radii of clusters, the thicknesses of UDGs only mildly increase with the decreasing $R/R_{200}$, consistent with the relatively high $p$ values from the K-S tests as shown in Table.~\ref{KS}. These results suggest a prominent environmental effect on morphological transformation of UDGs from outside to inside of $R_{200}$, but probably a weak environmental effect for the ones moving in clusters.

As discussed in section~\ref{sec:mass}, tidal interactions may be responsible to the puffed-up morphologies of UDGs in the denser environments; meanwhile, tidal interactions also extend sizes of UDGs \citep[e.g.,][]{Carleton19,Errani15}. In particular, UDGs embedded in cored halos expand much more significantly after tidal interactions, compared with the cuspy counterparts; however, the stellar mass within half-light radius is not significantly altered by the stripping process for the UDGs in cored halos \citep[cf.][]{Carleton19}. Since tidal influences are more prominent in higher-density environments (i.e., smaller $R/R_{200}$), we should therefore expect the sizes of stellar components to change with $R/R_{200}$, but luminosities to remain roughly constant. As shown by the $M_r$ and $r_{\rm{e}}$ (the median value and $1\sigma$ scatter are shown by the solid and dashed components, respectively) as functions of $R/R_{200}${\skip\footins=-\bigskipamount \footnote{Since Sample~3 applies a more conservative selection criterion of $r_{\rm{e,c}}=r_{\rm{e}}\sqrt{q}>1.5$~kpc (slightly different from the standard criterion $r_{\rm{e}}>1.5$~kpc for Samples~1 and 2), and Sample~3 UDGs are only located in $R<R_{200}$; therefore, in order to show the unbiased trends, we only show the radial trends for the low-$z$ UDGs with $r_{\rm{e,c}}>1.5$~kpc.}\skip\footins=\bigskipamount} in Fig.~\ref{dis_re_mag}, the sizes $r_{\rm{e}}$ for both of the high-mass and low-mass UDGs indeed slightly increase with decreasing $R/R_{200}$, while there is no obvious radial gradient in luminosity among our low-$z$ UDG samples.

We note that a large/dominant fraction of the present-day cluster UDGs might not be accreted as UDGs \citep{Ferre-Mateu18,Alabi18} but be transformed from typical dwarf progenitors under tidal interactions in clusters \citep[e.g.,][]{Jiang18,Liao19}; these `in-situ-transformed' UDGs may weaken the averaged radial trends of $r_{\rm{e}}$ and morphological transformation of UDGs.

Apart from the contamination of the `in-situ-transformed' cluster UDGs, the less efficient `environmental quenching' \citep[e.g., ram-pressure/tidal stripping or `strangulation';][]{Moore96,Larson80,Kawata08,Bekki09,Mayer07,Read05,Arraki14} may also reconcile the pronounced thickening of UDGs from $R>R_{200}$ to $R<R_{200}$, but mild thickening from $R\lesssim R_{200}$ to $R\sim 0$. In this scenario, the environmental quenching timescale may be relatively long \citep{Wheeler14}, so that the recently-accreted UDGs, primarily occupy the peripheries of clusters, may continue to retain their gas reservoirs, which cause galaxies to respond more impulsively to tides, significantly augmenting their morphological transformation \citep{Kazantzidis17}. However, most of the inner-region UDGs are devoid of gas, and thus their morphological transformation becomes less efficient, compared with the outer-region counterparts.

Another possibility is the contamination of interlopers. The UDG samples are selected based on their locations in the color-magnitude diagrams, and may include a fraction of reddened background massive star-forming interlopers. The contamination is more significant in the projected `outer-regions' of clusters where the clusters become less over-dense and comprising member galaxies with larger color scatters \citep[e.g.,][]{Lee16,Lee17,vanderBurg16}; therefore, these disky interlopers with small thicknesses, primarily located in the outer-regions of clusters, can introduce a more significant thickening of ``UDGs'' from $R>R_{200}$ to $R<R_{200}$. 

In addition, the mild radial trend in $R_{200}$ may be partly attributed to projection effects, i.e., the so-called `inner-region' ($R<0.5R_{200}$) UDG population may actually contains many projected middle/outer-region UDGs.

\subsection{UDG morphology evolution with redshifts}

\begin{figure}[!]
\centering
\includegraphics[width=\columnwidth]{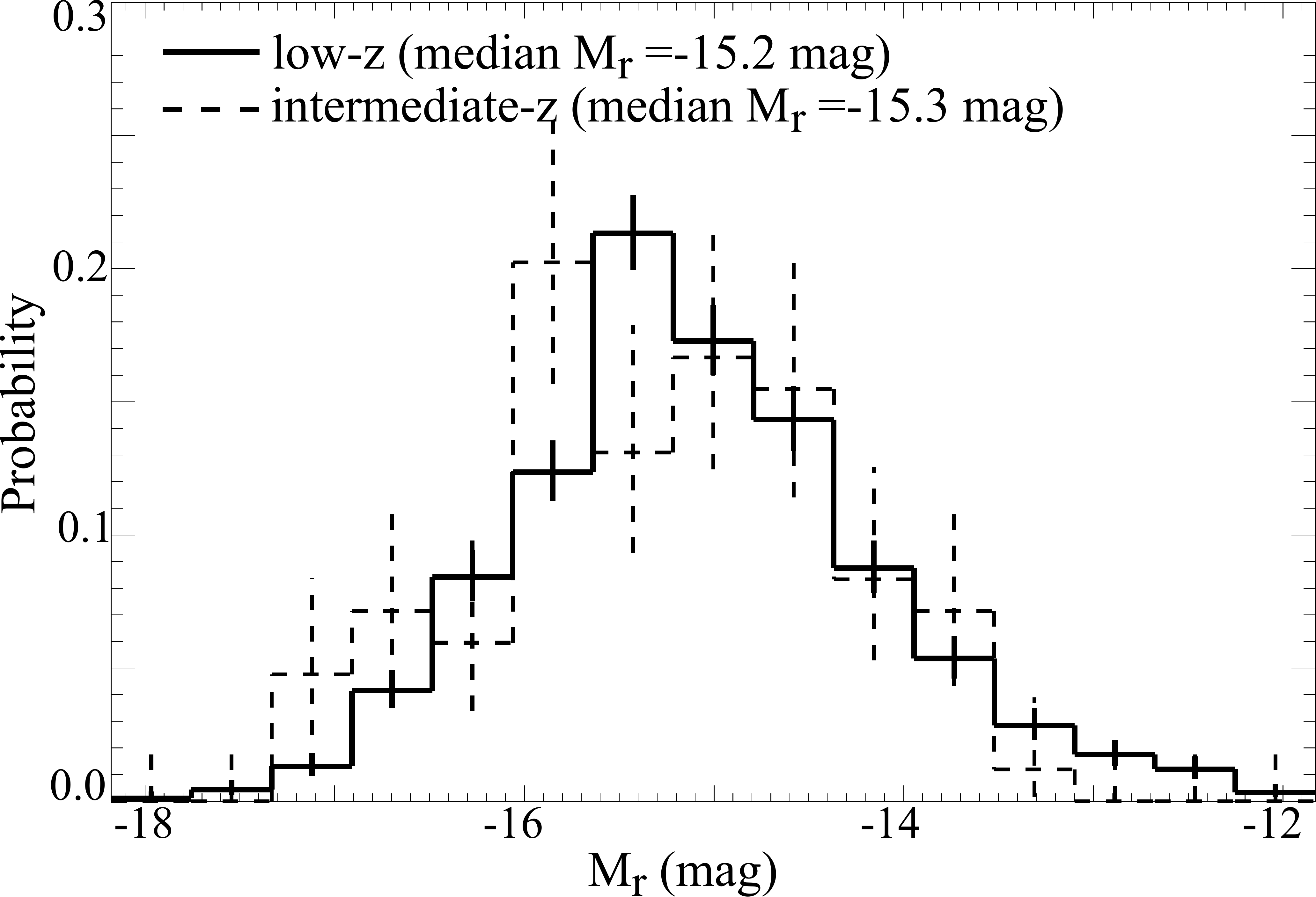}
\caption{The distributions of $M_r$ for the low-$z$ samples located in $R<R_{200}$ and intermediate-$z$ sample.}
\label{lowz_intz_mag}
\end{figure}

\begin{figure}[!]
\centering
\includegraphics[angle=-90,width=\columnwidth]{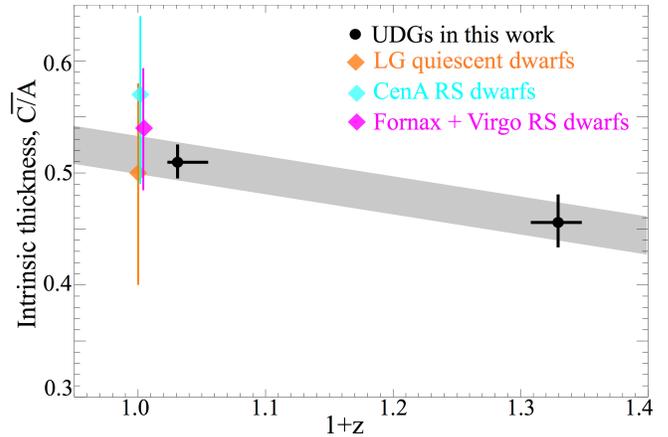}
\caption{The median intrinsic thickness $\bar{C/A}$ as a function of redshift for UDGs in this work (black), and dE/dSphs in the Local Group \citep[orange;][]{McConnachie12} and Centaurus~A galaxy group \citep[cyan;][]{Taylor17,Taylor18}, as well as dE/dSphs in the nearby clusters of Virgo and Fornax \citep[magenta;][]{Sanchez-Janssen19}. The gray shaded region shows the thickness trend for the UDGs located at the different redshifts. Analogous to Fig.~\ref{CA_location_mag}, the slope of the gray shaded trend is derived from linear fitting to the two black filled circles.}
\label{CA_redshift}
\end{figure}

The morphological transformation of UDGs from the intermediate to low redshift is also investigated. Since only $\sim7\%$ of the intermediate-$z$ UDGs are located beyond $R_{200}$, the intermediate-$z$ UDGs can be roughly treated as a cluster-UDG sample; besides, as shown in Fig.~\ref{lowz_intz_mag}, the luminosities of the intermediate-$z$ UDG sample and low-$z$ $R\leq R_{200}$ sample are approximately in the same $M_r$ range, with a median $M_r$ value of $-15.3$, and $-15.2$~mag, respectively. Therefore, the intermediate-$z$ UDG sample can be directly compared with the low-$z$ $R\leq R_{200}$ UDG sample. 

As explored in Fig.~\ref{CA_redshift}, the median intrinsic thickness of cluster UDGs slightly increases from intermediate to low redshift, probably suggesting that the cluster UDGs are marginally puffed up from $z\sim 0.35$ to 0. Therefore naively, we can suspect that the high-redshift, initial UDGs may be more flattened, and plausibly have a `disky' morphology. If we treat these intermediate-$z$ UDGs as the progenitors of the present-day cluster UDGs, it may imply that UDGs were originated from a formation mechanism \citep[e.g., high-spins of halos;][]{Amorisco16,Rong17} which can produce the `disky' morphologies in the first place.

Here, we also point out that the morphology difference between the high- and low-redshift UDGs may be driven by the small UDG sample from two intermediate-$z$ clusters, and thus the morphological transformation trend has large uncertainties and requires further confirmation with larger intermediate-$z$ UDG samples.

\section{Summary and Discussion}\label{sec:4}

With the data of apparent axis ratios for 1109 UDGs located in 17 low-$z$ ($z\sim 0.020\--0.063$) galaxy clusters and 84 UDGs in 2 intermediate-$z$ ($z\sim 0.308\--0.348$) clusters, we implement a Markov Chain Monte Carlo technology and assume a triaxial model to study the intrinsic morphologies of UDGs. In contrast to the conclusion of Burkert (2017), we emphasize that the UDG data favor the oblate-triaxial models over purely-prolate models.

The morphologies of UDGs are related to luminosity, environment, and redshift. For the low-$z$ UDGs, the ones with higher masses or located inside of the virial radii of clusters, are significantly puffed up, compared with the counterparts with low masses or located outside of the virial radii of clusters. Considering together the UDG morphologic dependences on luminosity and environment, we conclude that the most possible physical mechanism of leading to both of the two morphologic dependences is tidal interaction, essentially agreeing with the previous conclusion in, e.g., \cite{Conselice18}, though we cannot exclude that stellar feedback may also play a role on the UDG morphologic dependence on luminosity.

Note that the UDG thickness dependence on luminosity is distinct from that of the typical quiescent dE/dSphs in nearby clusters and groups, suggesting that tidal interaction plays a much more crucial role in the evolution of UDGs compared with the typical dwarfs. It probably implies that UDGs may not be simply treated as an extension of the dE/dSph class with similar evolutionary histories; they may differ not only in size.

From intermediate to low redshift, the morphologies of cluster UDGs become marginally puffier, and have broader ranges of ellipticity and triaxiality, plausibly suggesting a formation mechanism producing `disky' morphologies for the high-redshift, newly-born UDG progenitors, e.g., being formed in the high-specific-angular-momentum halos \citep{Amorisco16,Rong17}. However, the number of UDGs at relatively high redshifts is small; we strongly encourage further high-spatial-resolution, deep surveys for high-redshift UDGs to examine this conclusion.

Note that, tidal interactions can alleviate the tension between the detected high-specific-angular-momenta of the field UDGs \citep{Leisman17} and large dispersion or no signs for rotation in cluster-UDGs \citep[e.g.,][]{vanDokkum16,Danieli19,Chilingarian19}: the high-redshift, rotationally-supported, field UDGs (resembling present-day dwarf irregulars) might be originated in the halos with high specific angular momenta, and have initial shapes with relatively small thicknesses; when UDGs were accreted to high-density environments, the tidal interactions can efficiently reduce their angular momenta and transform UDG morphologies, to make their two-dimensional morphologies resemble dEs, dS0s, and dSphs \citep{Ferre-Mateu18,Alabi18,Vijayaraghavan13,Vijayaraghavan15,DeLucia12,Rong15b,Rong16,Cortese06,Zabludoff98,McGee09,Peng12}. 

\acknowledgments

We thank the referee for her/his helpful comments and suggestion. We acknowledge the publicly available UDG catalogs in \cite{Yagi16} and \cite{Lee17}. Y.R. acknowledges the helpful comments and suggestions from M. G. Lee, R. S\'anchez-Janssen, L. Gao, Q. Guo, S.-H. Liao, and J. Wang, as well as funding supports from FONDECYT Postdoctoral Fellowship Project No.~3190354 and NSFC grant No.\,11703037. T.H.P. acknowledges support through FONDECYT Regular project 1161817 and CONICYT project Basal AFB-170002. H.X.Z. acknowledges support from the CAS Pioneer Hundred Talents Program and the NSFC grant 11421303. This work is also supported by CAS South America Center for Astronomy (CASSACA), Chinese Academy of Sciences (CAS).

This research has made use of the NASA Astrophysics Data System Bibliographic Services, the NASA Extragalactic Database, {\sc Python/Emcee} v.2.2.1 \citep[][\url{https://emcee.readthedocs.io/en/v2.2.1/}]{Foreman-Mackey13} package, {\textsc{GALFIT}} \citep{Peng02,Peng10}, as well as ds9 (a tool for data visualization supported by the Chandra X-ray Science Center (CXC) and the High Energy Astrophysics Science Archive Center (HEASARC) with support from the JWST Mission office at the Space Telescope Science Institute for 3D visualization). We also acknowledge the related literature of \cite{Goodman10,Jones01,VanderPlas12,Rong18b,Rong17b,ast13,Hunter07}.

\vspace{5mm}

\appendix
\counterwithin{figure}{section}

\section{MCMC results versus observations}\label{sec:aa}

In this section, we test whether the MCMC results are robust and can well recover the observed $q$ distributions of UDGs. Particularly for the low-$z$ samples, the $\bar{T}$ posterior pdfs actually contain two peaks (local maximum likelihood) overlapping at about $\bar{T}\simeq 0.8$ (cf. Figs.~\ref{MC_Yagi}, \ref{MC_Pina}, and \ref{MC_Burg8}; the primary peak at $\bar{T}\sim 0.3\--0.4$ and secondary peak at $\bar{T}\sim 1.0$), and the posterior pdfs of $\bar{T}<0.8$ resembling Gaussian distribution correspond to the triaxial models, while posterior pdfs of $\bar{T}>0.8$ suggest the nearly-prolate models. 

It is worth to note that, in MCMC, these $\bar{T}>0.8$ steps are not `burn-in' steps which should be discarded. Indeed, the MCMC chain converges to an equilibrium distribution after approximately 200 `burn-in' steps \citep{Sanchez-Janssen16} and we actually apply `nburn'=400 in the affine-invariant MCMC algorithm implemented in the Python {\textsc{emcee}} package \citep{Foreman-Mackey13}. We also set `nburn'=500, 1000, 2000, and 4000 (remain the other parameters unchanged), and find that the $\bar{T}$ posterior pdfs of the low-$z$ samples always show double-peaks. 

\begin{figure*}[!t]
\centering
\includegraphics[width=0.8\textwidth]{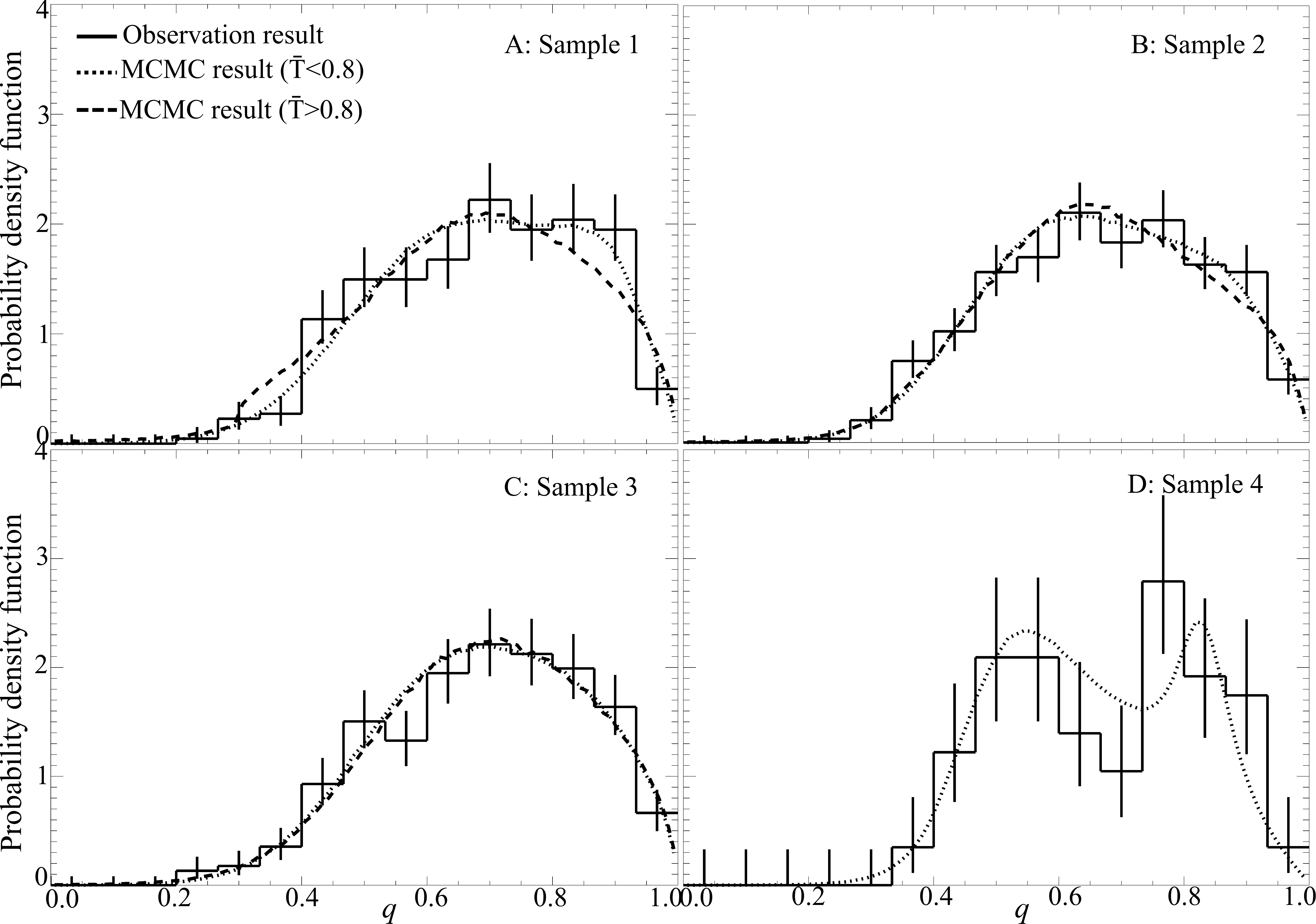}
\caption{The comparison between the observational $q$ distributions (solid histograms) and recovered distributions by the triaxial models (with the sets of ($\bar{E}$, $\sigma_E$, $\bar{T}$, $\sigma_T$) corresponding to $\bar{T}<0.8$; dotted) and purely-prolate models (with the sets of ($\bar{E}$, $\sigma_E$, $\bar{T}$, $\sigma_T$) corresponding to $\bar{T}>0.8$; dashed), derived from MCMC, for the four UDG samples.}
\label{comparison}
\end{figure*}

Further, as shown in Fig.~\ref{comparison}, we find that both of the triaxial models and nearly prolate models generated from MCMC can well recover the observed $q$ distributions{\footnote{Actually, compared with the nearly prolate models, the triaxial models can reproduce the marginally flatter $q$ distributions in the range of $q\sim 0.6\--0.9$, more closely resembling the observed $q$ distributions, as shown in Fig.~\ref{comparison}}}. In summary, the MCMC results are robust and the $\bar{T}>0.8$ steps are not `burn-in' steps.

\section{UDG morphologic dependence on luminosity}\label{sec:bb}

\begin{figure*}[!b]
\centering
\includegraphics[scale=0.2]{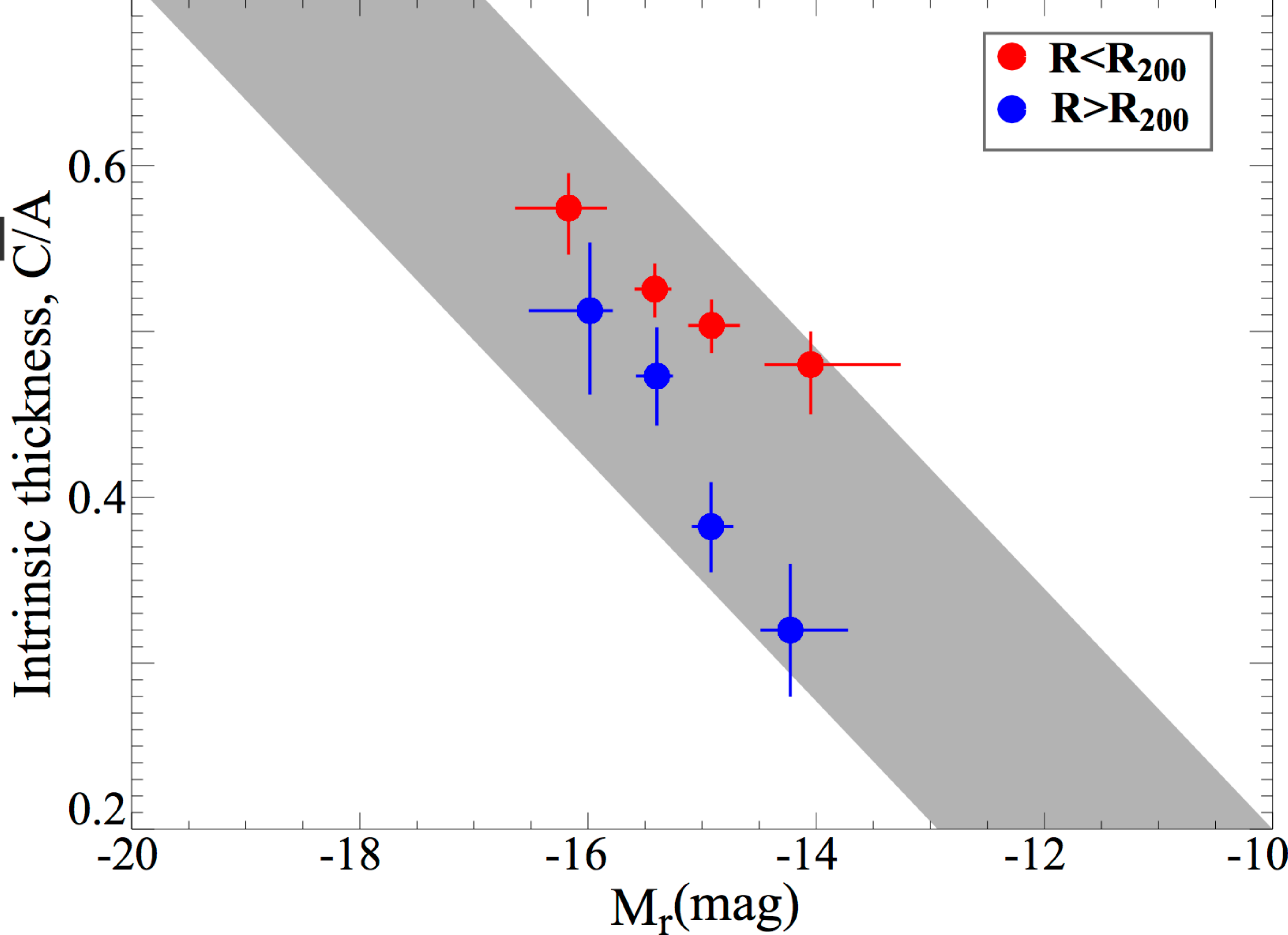}
\caption{The thickness of UDGs as a function of absolute magnitudes. The red and blue filled circles denote the UDGs in and out of $R_{200}$, respectively. Analgous to Fig.~\ref{CA_location_mag}, the grey shaded region shows the UDG thickness trend that the more-luminous UDGs show puffier morphologies.}
\label{DOL}
\end{figure*}

In this section, for the low-$z$ UDGs, we split both of the $R_{200}$ and $R>R_{200}$ samples into four subsamples (instead of the two subsamples with $M_r<-15.2$~mag and $M_r>-15.2$~mag) according to their absolute magnitudes, to re-test the UDG thickness dependence on luminosity. As shown in Fig.~\ref{DOL}, we can still find the significant decreasing trend of UDG thicknesses with the decreasing luminosities, though the uncertainties of thicknesses of UDGs are relatively large.

\vspace{5mm}
\clearpage


\end{document}